\documentclass[jgrga]{agutex}

\usepackage{amsmath,amssymb,alltt,verbatim,xspace}

\newcommand{\url}[1]{\texttt{#1}}
\newcommand{\href}[2]{\texttt{#2}}

\ifx\pdftexversion\undefined
  \usepackage[final,dvips]{graphicx}
\else
  \usepackage[final,pdftex]{graphicx}
\fi

\newcommand\bb{\mathbf{b}}

\newcommand\bQ{\mathbf{Q}}
\newcommand\bu{\mathbf{u}}
\newcommand\bv{\mathbf{v}}
\newcommand\bU{\mathbf{U}}

\newcommand{\ddt}[1]{\ensuremath{\frac{\partial #1}{\partial t}}}
\newcommand{\ddx}[1]{\ensuremath{\frac{\partial #1}{\partial x}}}

\newcommand{\ddz}[1]{\ensuremath{\frac{\partial #1}{\partial z}}}

\newcommand{\Div}{\nabla\cdot}
\newcommand\eps{\epsilon}
\newcommand{\grad}{\nabla}

\newcommand{\pp}[2]{\ensuremath{\frac{\partial #1}{\partial #2}}}
\newcommand{\ppp}[2]{\ensuremath{\frac{\partial^2 #1}{\partial #2^2}}}

\newcommand\Hmelt{W}

\authorrunninghead{BUELER AND BROWN}

\titlerunninghead{SHALLOW SHELF APPROXIMATION AS A ``SLIDING LAW''}

\authoraddr{Ed Bueler,
Department of Mathematics and Statistics, Box 756660,
University of Alaska Fairbanks,
Fairbanks, AK 99775-6660, USA. (ffelb@uaf.edu)}

\begin{document}

\title{The shallow shelf approximation as a ``sliding law'' \\ in a thermomechanically coupled ice sheet model}

\author{Ed Bueler [DRAFT AS SUBMITTED; \today]}
\affil{Department of Mathematics and Statistics, 
       University of Alaska, Fairbanks, Alaska, USA}

\author{Jed Brown}
\affil{VAW A-15, ETH Zentrum, 8092 Zurich, Switzerland}

\date{\today}

\begin{abstract}
The shallow shelf approximation is a better ``sliding law'' for ice sheet modeling than those sliding laws in which basal velocity is a function of driving stress.  The shallow shelf approximation as formulated by \emph{Schoof} [2006a] is well-suited to this use.  Our new thermomechanically coupled sliding scheme is based on a plasticity assumption about the strength of the saturated till underlying the ice sheet in which the till yield stress is given by a Mohr-Coulomb formula using a modeled pore water pressure.  Using this scheme, our prognostic whole ice sheet model has convincing ice streams.  Driving stress is balanced in part by membrane stresses, the model is computable at high spatial resolution in parallel, it is stable with respect to parameter changes, and it produces surface velocities seen in actual ice streams.
\end{abstract}

\begin{article}
\section{Introduction}\label{sect:intro}

A well-known difficulty with numerical ice sheet models is their inability to model the large range of ice flow speeds observed in real ice sheets \citep{ShepardWingham,TrufferFahnestock,VaughanArthern}.  Observed surface speeds for ice flow in the Greenland ice sheet, for example, range from less than $10$ meters per year in large areas of the interior \citep[compare][]{Greve97Greenland,JoughinetalGrBal97} to more than $10$ km per year in three outlet glaciers  \citep{HowatJoughinScambos,JoughinAbdalatiFahnestock}.  Existing Greenland ice sheet models have not, however, reported (published) ice surface speeds in excess of $100$ m per year \citep{Greve97Greenland,Greve00, RitzFabreLetreguilly,SaitoAbeOuchi, TarasovPeltier}.

Fast grounded ice flow, in ice streams and outlet glaciers to differing degrees \citep{TrufferEchelmeyer}, arises from some combination of sliding, over a rigid or deformable mineral bed, and shear deformation of the lowest wet, dirty layers of ice.  Unfortunately and fundamentally, however, remote sensing provides no high quality spatially-distributed observations of conditions at or near the ice base with which to constrain models of fast flow.  There is a triple need to improve observations, to use existing surface observations more effectively, and to improve models of ice flow including sliding.

This paper approaches modeling fast ice stream motion pragmatically, within the high-resolution, comprehensive, thermomechanically-coupled, and time-dependent Parallel Ice Sheet Model \citep[``PISM'';][]{pism-user-manual}.  The basal mechanical model we add here is based on a spatially-distributed till friction angle \citep{Paterson}.  We demonstrate that our model responds in a reasonable way to changes in till friction angle and other major parameter choices including grid refinement.  We believe that the model is a credible model of shallow ice streams.  In our model, ice sheet geometry and thermodynamical fields (ice temperature and effective thickness of basal water) evolve together within a unified shallow framework.

\begin{figure}
\noindent\includegraphics[width=20pc]{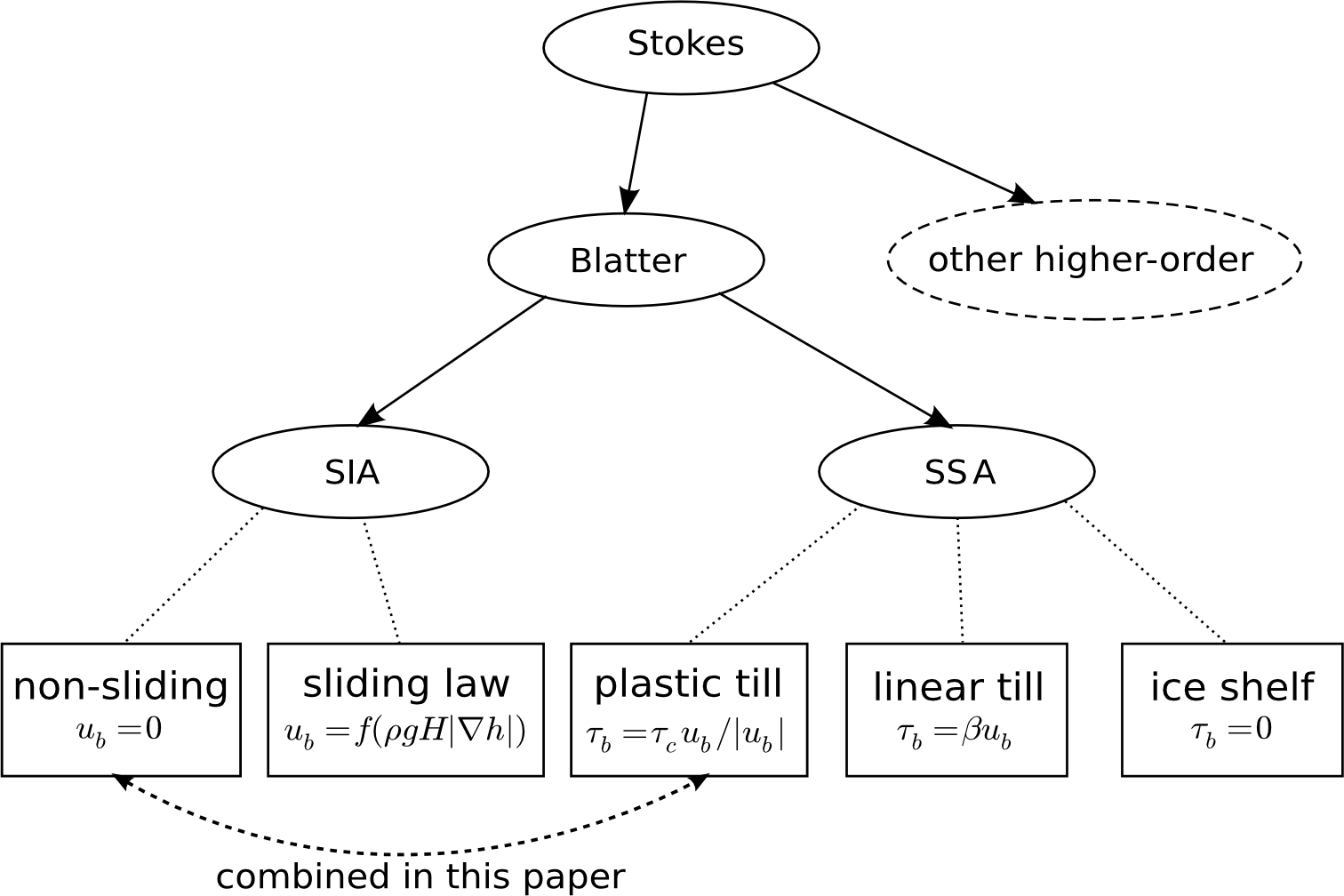}
\caption{A hierarchy of ice dynamics models (ellipses) with sliding parameterizations which have been applieds to the shallower models (boxes).  Solid arrows show rigorous small-parameter shallowness arguments, while $u_b$ denote basal ice velocity and $\tau_b$ basal shear stress.}
\label{fig:hierarchy}
\end{figure}

Fast-flowing simulated ice is not useful in a model if it arises from unreasonable physics.  All ice sheet models incorporate approximations, and most models, including ours, use the actual shallowness of ice sheets to simplify the equations and reduce computational cost.  There are choices in parameterizing the sliding, however.  We recall some continuum flow models applied to ice sheets and sliding in the hierarchy in Figure \ref{fig:hierarchy}.  All the illustrated models describe ice as a slow, non-linearly viscous, isotropic fluid, though these qualities are approximations too.

The simplest and shallowest models are called the \emph{shallow ice approximation} (SIA) \citep{Hutter,MorlandJohnson} and the \emph{shallow shelf approximation} \citep[``SSA'';][]{Morland,WeisGreveHutter}.  Rigorous small-parameter arguments explain how to simplify from Stokes to ``higher-order models''  \citep{Blatter,Hindmarsh04compare}, and from the \cite{Blatter} model to the SIA and SSA \citep{SchoofHindmarsh}.

Thermomechanically coupled, shallow, grounded, and non-sliding (frozen) base ice sheet models based upon the SIA are relatively well understood \citep{BBL,EISMINT00}.  Large portions of actual ice sheets have bases which experience minimal sliding and have modest bed topography.  For those parts the nonsliding SIA is a second order \citep{Fowler} theory which predicts a reasonable distribution of flow at rates which compare well to observations (e.g.~\citep{Greve97Greenland}).

As noted, faster ice flow is a combination of sliding over the mineral base along with deformation of an ice-and-till layer at the base.  We necessarily lump these mechanisms as ``sliding'' in the language of this paper, because of the lack of observational techniques necessary to distinguish mechanisms at whole ice sheet scale.  In any case, sliding applies a boundary force (stress) to the base of the ice mass, the effect of which is distributed by the stress balance.  In our discussion of stress balance models below, components of the stress tensor will be separated into shear in planes parallel to the geoid (``horizontal plane shear'') versus the other ``membrane'' stresses \citep{Hindmarsh04compare,HindmarshMembrane}.

The SIA has no mechanism for balancing the driving stress partially by membrane stresses, nor any method of incorporating the sliding stress into a stress balance at all.  Nonetheless ``sliding laws'' have been added to SIA models anyway \citep[among others]{Greve97Greenland,GreveTakahamaCalov,HuybrechtsdeWolde,EISMINT00}.  Such laws necessarily describe the velocity of the base of the ice as a function of the driving stress, namely the product of the cryostatic pressure at the base times the surface slope \citep{Paterson}.  Furthermore, sliding is usually assumed to be insignificant when the ice base is frozen, but to ``turn on'' only when the ice base reaches the pressure-melting temperature \citep[e.g.][]{GreveTakahamaCalov,EISMINT00} or to change discontinuously at sufficient depth below sea level \citep{HuybrechtsdeWolde,TarasovPeltier}.  Unfortunately, the underlying SIA continuum model then propagates jump discontinuities in the horizontal velocity field through the entire ice column.  In turn, this produces unbounded vertical velocities because of the incompressibility of the ice \citep{Fowler01}.  These facts about the continuum model have the unfortunate consequence that numerical schemes for temperature or age will not converge under grid refinement, and, in fact, flow predictions from such numerical models become more unreasonable on finer grids (Appendix B).

It remains critical, however, that modeled sliding depends on the ice temperature and/or on the amount of liquid water present at the ice base.  Furthermore these quantities (temperature and basal water) must evolve if modeled ice streams are to exhibit the observed energy-balance-dependent behavior actually seen \citep{Raymondenergy,SchoofMargins}.  The coupling could potentially include a detailed model of till deformation \citep{PollardDeConto}, and perhaps also a distributed model for melt water conservation \citep{JohnsonFastook}, but the model in this paper is simpler.  An effective thickness for the water stored in the till is computed by time-integrating the rate of melt water production at the ice base.  A negative melt rate (freeze-on) is allowed.  The estimated pore water pressure is function of this effective thickness.  The pore water pressure is used when computing the effective pressure on the mineral till.  The Mohr-Coulomb criterion then describes the yield stress of the saturated till as the product of the effective pressure on the till and the tangent of the till friction angle \citep{Paterson}.  Till strength (yield stress) then enters as a term in a membrane stress balance, so the sliding velocity depends non-locally on till strength.

Large ice shelves, with zero till strength, are well-described by the SSA model \citep{MacAyealetal,Morland,WeisGreveHutter}.  Some published Antarctic ice sheet models use the SSA for the force balance in ice shelves \citep{Huybrechts90,HuybrechtsdeWolde,Ritzetal2001}.  Furthermore, diagnostic models based on the ``dragging ice shelf'' extension of the SSA have been applied to individual ice streams or ice stream basins \citep{HulbeMacAyeal,MacAyeal}.  These models have been exploited to recover the mechanical properties of the sliding ice base from observed surface velocities \citep[``inverse modeling'';][]{Joughinetal2001,JoughinMacAyealTulaczyk}.  Though \cite{MacAyeal} proposed a linear viscous till deformation model, the mechanical properties of the till in real ice streams and outlet glaciers is a largely open question, with credible arguments for linear and plastic (Coulomb) models \citep{JoughinMacAyealTulaczyk,SchoofTill}.  Power law formulations in between these extremes have also been considered \citep{SchoofCoulombBlatter}.

The current paper assumes that the till material behaves plasticly.  This is because the best-understood continuum model which includes the SSA stress balance uses plastic till as part of a well-posed free boundary problem for the velocity.  In fact, \cite{SchoofStream} has recognized, in the isothermal and time-\emph{independent} case, that including plastic failure of the basal till under flowing ice gives emergent ice streams within a whole ice sheet system.  Our inclusion of Schoof's model into a time-\emph{dependent} model is new.  The locations of sliding flow are not predetermined in Schoof's model, and ours.  In ours these locations also evolve.

We use the SSA model \emph{as a sliding law} in a thermomechanically-coupled SIA model.  The boundaries of the sliding regions are locations where the sliding velocities go to zero in a well-behaved way \citep{SchoofStream}, so the problems described in Appendix B do not arise.  The majority of the flow, by map-plane area or ice volume fraction, is by horizontal plane shear according to the non-sliding SIA.  There are, however, ice streams which flow partly by the SIA but mostly by additional basal sliding constrained by the SSA balance of membrane stresses.  The ice is allowed to slide anywhere, but sliding does not occur in the majority of the basal area because the till is too strong.  Though till friction angle is time-independent, the actual till strength (yield stress) evolves in a thermomechanically coupled and time-dependent manner (through evolution of the modeled pore water pressure).

The sliding velocity field arising at a particular time from the solution of the SSA is averaged with the velocity field which solves the nonsliding SIA at that time.  This is the sense in which the SSA is a sliding law.  As a result, the velocity field is a weighted average of two published ice sheet models, namely the nonsliding SIA \citep[etc.]{Hutter} and a plastic till form of the SSA \citep{SchoofStream}.  The average is weighted toward the SIA if sliding is slow and toward the SSA if sliding is fast (subsection \ref{subsect:super}).  

As suggested by Figure \ref{fig:hierarchy}, our stress balance combination model is not as complete as a number of ``higher-order'' or ``full Stokes'' (ISMIP-HOM) alternatives.  The stress balance for the Stokes model is, however, a three independent variable problem for the three components of the velocity field, at each time step.  It has currently only been solved either for thermomechanically coupled but diagnostic problems on glaciers \citep[e.g.][]{Zwingeretal07} or with geometry evolution but isothermally and again on glacier-sized problems \citep[e.g.][]{HOMtcd}.  The stress balance for the \cite{Blatter} model is likewise a three independent variable problem at each time step which has been implemented on larger systems but not at high resolution \citep{Pattyn03,SaitoEISMINT}.  Furthermore these higher order models have not demonstrated clear success for sliding flow even in flowline glacier modeling circumstances \citep{HOMtcd}.

Our model is time-dependent in the usual sense that ice sheet geometry evolves according to a mass continuity equation.  The flow is thermomechanically coupled in the usual senses that the temperature affects the softness of the ice and there is dissipation heating from ice deformation and frictional heating from sliding.

\section{Continuum model}\label{sect:model}

Our description of the continuum model is organized into seven subsections: \emph{mass continuity}, \emph{flow law}, \emph{conservation of energy}, \emph{basal melt}, \emph{SIA version of the stress balance}, \emph{SSA version of the stress balance}, \emph{basal mechanics}, and \emph{combination of velocity fields}.  Notation follows Table \ref{tab:notation}.

The time-independent boundary data are bed elevation $b(x_1,x_2)$, ice surface temperature $T_s(x_1,x_2)$, accumulation rate $M(x_1,x_2)$.  The modeled time-dependent unknowns are the ice thickness $H(t,x_1,x_2)$, surface elevation $h(t,x_1,x_2)$, basal melt rate $S(t,x_1,x_2)$, thickness of a melt water layer $\Hmelt(t,x_1,x_2)$, temperature $T(t,x_1,x_2,x_3)$, and vector velocity $\bU(t,x_1,x_2,x_3)$.  The relation $h=H+b$ holds always because only grounded ice sheets are considered here.  Let $D_{ij}$ be the strain rate tensor, that is, $D_{ij} = (1/2) \left(\partial U_i/\partial x_j + \partial U_j/\partial x_i\right)$, $\sigma_{ij}$ the full (Cauchy) stress tensor, $p = -(1/3)\left(\sigma_{11}+\sigma_{22}+\sigma_{33}\right)$ the pressure, and $\tau_{ij}=\sigma_{ij}+p\delta_{ij}$ the deviatoric stress tensor.

\subsection{Mass continuity}  The upper surface $z=h$ of the ice is a free surface.  Let $U_{h,i} = U_i(z\!\!=\!\!h)$ be the surface value of the velocity.  Then, using a small slope assumption for the ice surface \citep{Fowler},
\begin{equation}\label{surfkine}
   \ddt{h} = M + U_{h,3} - \grad h \cdot (U_{h,1},U_{h,2})
\end{equation}
is the surface kinematical equation; $\grad$ is in the horizontal variables.  A similar equation applies at the base of the ice:
\begin{equation}\label{basekine}
   0 = S + U_{b,3} - \grad b \cdot (U_{b,1},U_{b,2}).
\end{equation}

Define $\bQ=\bar \bU H$, the horizontal ice flux, where $\bar \bU$ is the vertically-averaged horizontal velocity.  (For the non-sliding SIA (subsection \ref{subsect:sia} below), the expression $\bQ = - D_{\text{SIA}}\;\grad h$ also applies, where $D_{\text{SIA}}$ is a positive diffusivity \citep{BBL}.  For membrane stress balanced sliding, a diffusivity can not be meaningfully defined because the ice flux $\bQ$ is not generally parallel to $\grad h$.  Instead we will treat the mass continuity problem associated to basal sliding as a generic transport problem (subsection \ref{subsect:massconnumerics} below).)

Ice is an incompressible fluid, so $D_{11} +D_{22} + D_{33} = 0$.  The equation of mass continuity
\begin{equation}\label{mapplanecont}
   \ddt{H} = M - S - \Div \bQ.
\end{equation}
can be derived from incompressibility by using the ice surface and base kinematic equations.  We will solve Equation \eqref{mapplanecont} numerically to compute the ice thickness at each grid point at each time step.

The vertical velocity within the ice is, by incompressibility,
\begin{equation}\label{vertvel}
   U_3 = - S + \grad b \cdot (U_{b,1},U_{b,2}) - \int_b^{x_3} \Div (U_1,U_2)\,d\zeta
\end{equation}
We have included the (ice-equivalent) basal melt rate $S$ into the vertical velocity of the ice.  In particular, the basal melt rate influences the vertical advection term in conservation of energy (subsection \ref{subsect:conserve}, next).  Also, we have included the contribution of the basal melt rate $S$ to the mass continuity equation \eqref{mapplanecont}; compare \citep{EISMINT00}.  The basal melt rate will also feed back to affect ice dynamics by influencing the till yield stress (subsection \ref{subsect:tillmech}).

\subsection{Flow law for ice}\label{subsect:flowlaw}  Ice is modeled as a nonlinearly viscous isotropic fluid with a constitutive relation of Arrhenius-Glen-Nye form \citep{Paterson}
\begin{equation}\label{constitutive}
  D_{ij} = A(T^*) \tau^{n-1} \tau_{ij}
\end{equation}
where $T^* = T + \beta (h-z)$ is the pressure-corrected temperature, $n=3$, and $2 \tau^2 = \tau_{ij}\tau_{ij}$ determines the second stress invariant $\tau$ \citep{Fowler}.  This form of the flow law is used of the SIA stress balance (subsection \ref{subsect:sia}).  The ice softness (flow factor) $A(T^*)$ is determined by the formulas of \cite{PatersonBudd}:
	$$A(T^*) = \begin{cases} (3.61 \times 10^{-13})\,\, e^{-6.0\times 10^4 / (R T^*)}, & T^* \le 263.15\,\text{K}, \\ (1.73 \times 10^{+3})\,\, e^{-13.9\times 10^4 / (R T^*)}, & T^* > 263.15\,\text{K}. \end{cases}$$

For the SSA stress balance (subsection \ref{subsect:ssa}) we state the flow law in viscosity form
\begin{equation}\label{viscosityform}
   \tau_{ij} = 2\, \nu(T^*,D)\,D_{ij}.
\end{equation}
In this case the nonlinear viscosity $\nu$ satisfies $2 \nu(T^*,D) = B(T^*) D^{(1/n)-1}$ where $2 D^2 = D_{ij}D_{ij}$ and $B(T^*)=A(T^*)^{-1/n}$ is the ice hardness.

\subsection{Conservation of energy}\label{subsect:conserve}  We use a standard conservation of energy model for cold ice \citep[not polythermal;][]{Greve}.  Shallowness, namely a small ratio $\eps=[H]/[l]$ of typical ice thickness $[H]$ to typical ice sheet width $[l]$, simplifies the equation so that the horizontal conduction terms drop out.  This simplification occurs both in the shallowness argument that leads to the SIA \citep{Fowler} and the one that leads to the SSA \citep{WeisGreveHutter}, so we adopt it throughout:
\begin{equation}\label{conserve}
   \rho_i c_i \left(\pp{T}{t} + U_1 \pp{T}{x_1} + U_2 \pp{T}{x_2} + U_3 \pp{T}{z}\right)  = k_i\ppp{T}{z} + \Sigma
\end{equation}
Here $\rho_i,c_i,k_i$ are constant material properties of the ice (see Table \ref{tab:notation}).  The term in parentheses on the left of \eqref{conserve} is the material derivative $dT/dt$.  We denote by $\Sigma$ the rate at which straining is converted to heat and is applied to warming the ice volumetrically.

In fact, recalling that ice sheets are a slow flow, all work done by the driving stress of gravity is immediately deposited as heat into the ice or is used for producing melt water at the base (subsection \ref{subsect:tillmech}).  In the case of the full Stokes model,
\begin{equation}\label{SigmaFS}
\Sigma_{\text{FS}} = \sum_{i,j=1}^3 D_{ij} \tau_{ij} = 2 B(T^*)D^{(1/n)+1}
\end{equation}
\citep{Greve}.  We compute the strain heating $\Sigma$ in equation \eqref{conserve} using the shallow approximations of $\Sigma_{\text{FS}}$ described in subsection \ref{subsect:super}.

Frictional heating occurs at the ice-lithosphere interface where the ice is sliding.  The rate of heating is $-\tau_b \cdot \bU_b$, a positive value with SI units of $\text{J}\, \text{m}^{-2}\, \text{s}^{-1}$ \citep[e.g., chapter 10 of][]{Paterson}.  This rate times the area of the horizontal face of a grid cell times the length of a time step is an amount of heat which is added at each time step to the grid cell at the ice-lithosphere interface.

Lithosphere ``thermal inertia'' is an important physical process in the thermomechanical regulation of ice stream flow, and it is a standard part of paleo ice sheet modeling (e.g.~\citep{Greve97Greenland,RitzFabreLetreguilly}).  We track the temperature of the lithosphere through a simplified conservation of energy model which again lacks horizontal conduction terms because of shallowness:
\begin{equation}\label{conserveLitho}
   \pp{T}{t} = K_r\ppp{T}{z}.
\end{equation}
Here $k_r = 3.0\,\text{W}\,(\text{K}\;\text{m})^{-1}$, $\rho_r = 3300\,\text{kg}\,\text{m}^{-3}$, and $c_r = 2009\,\text{J}\, (\text{kg}\;\text{K})^{-1}$, so $K_r = k_r / (\rho_r c_r) = 9.09 \times 10^{-7}\,\text{m}^2\,\text{s}^{-1}$.  The thermal model extends downwards a distance of $D_L=515$ m into the lithosphere.  This depth $D_L$ is chosen for convenience because no theory known to the authors identifies a preferred value.  We observe that ice stream dynamics in our model are significantly different if $D_L=0$, becayse with no lithosphere thermal model there is more oscillation in the streaming flow, including more frequent complete halts of the flow.

The boundary conditions for Equation \eqref{conserve} apply a surface temperature to the top of each ice column and a geothermal flux to the base of the lithosphere layer,
	$$T\big|_{(z=h)} = T_s(x_1,x_2), \qquad - k \pp{T}{x_3}\Big|_{(z = b - D_L)} = G_0.$$

\subsection{Basal melt}\label{subsect:melt}  We compute a basal melt rate, and model the storage of melted water locally at the base of the ice column, as follows:  If at some time step the ice in a grid cell reaches the pressure melting temperature $T_0^*=T_0 - \beta(h-z)$ then our model converts a small fraction of the excess heat produced during that time step to melt water which appears at the base of the ice column.  This fraction depends on the height above the bed.  Concretely, for a time step $\Delta t$, ``cold ice'' Equation \eqref{conserve} predicts tentative temperature $\tilde T(t+\Delta t)$, which might actually be above the pressure melting temperature.  Our basal melt rate is
\begin{equation}\label{basalmeltrate}
S \Delta t = \frac{c_p}{L} \int_b^{b+100} \left(\tilde T(t+\Delta t) - T_0^*\right)\, \left[0.2\,\frac{b+100-z}{100}\right]\,dz.
\end{equation}
The unitless number in square parentheses is the fraction of the melt water produced at height $z$ which appears as free liquid water at the base.  This fraction decreases linearly from 20\% at the base to 0\% at 100 meters above the base, and is zero above that.  The basal melt rate $S$ can be negative in the case where water is freezing on to the bottom of the ice column.  Freeze-on only occurs if stored water is available at the base; storage is addressed next.

The water produced by melting of ice near the base is locally stored in till.  This water is described by an effective thickness $\Hmelt$.  It is available for weakening the till mechanically and for refreezing.  Thickness $\Hmelt$ is updated by adding the basal melt rate $S$ at each time step, but with additional spatial diffusion:
\begin{equation}\label{Hmelt}
   \pp{\Hmelt}{t} = S + K_{\text{melt}}\left(\ppp{\Hmelt}{x_1} + \ppp{\Hmelt}{x_2}\right),
\end{equation}
where $K_{\text{melt}} = \tilde L^2 / (2 \tilde t^2)$, $\tilde L = 20$ km, and $\tilde t = 1000$ years.  That is, the melt water is produced locally in each column according to \eqref{basalmeltrate} but it diffuses so that a delta mass spreads to a normal distribution with standard deviation of $20$ km in $1000$ years.  Finally, $\Hmelt$ is limited to at least zero and at most $W_0=2$ meters.  The upper limit forms a minimal model for basal drainage.

Once the basal melt rate $S$ and the stored basal water thickness $\Hmelt$ are computed for time $t+\Delta t$, all temperatures within the ice are cut off at the pressure melting temperature $T_0^*$: $T(t+\Delta t) = \min\{T_0^*,\tilde T(t+\Delta t)\}$.

A front-tracking \citep{Greve} or enthalpy gradient \citep{AschwandenBlatter} polythermal model will improve the above melt-water production model.

Subsection \ref{subsect:tillmech} addresses how $W$ is used to compute till yield stress.  The connection between stored basal water and till yield stress is a critical coupling in our model.

\subsection{SIA version of the stress balance}\label{subsect:sia}  The fundamental stress balance 
of glaciology is the Stokes model for a slow flowing fluid \citep{Fowler}.  Here ``slow'' has the precise meaning that inertia can be neglected in the force (stress) balance.  In this paper we combine two shallow versions of the Stokes model.  One is the shallow ice approximation (SIA) \citep{Hutter,MorlandJohnson}.  We use only the well-justified non-sliding version of the SIA, a non-Newtonian lubrication approximation for ice \citep{Fowler}.   The nonsliding SIA gives an ice sheet flow in which the driving force of gravity is balanced exclusively by horizontal plane shear.

Denoting the velocity field which solves this stress balance by $\bu=(u_1,u_2,u_3)$, the SIA can be written as a single equation
\begin{equation}\label{balanceSIA}
   \left(\pp{u_1}{z},\pp{u_2}{z}\right) = - 2 (\rho g)^n\,A(T^*)\,(h-z)^n\,|\grad h|^{n-1}\,\grad h.
\end{equation}
Equation \eqref{balanceSIA} computes shear strain rates $D_{13},D_{23}$ from local information, in the map-plane sense.  Vertical integration gives
\begin{equation}\label{velocitySIA}
   \left(u_1,u_2\right) = - 2 (\rho g)^n\,|\grad h|^{n-1}\,\left[\int_b^z A(T^*)\,(h-\zeta)^n\,d\zeta\right]\,\grad h.
\end{equation}
In our model the integral in square brackets is computed numerically by the trapezoid rule on a vertical grid which is finest near the base of the ice.  

Throughout this paper the pressure is modeled as cyrostatic, $p = \rho g (h-z)$.
This approximation holds in both the SIA \citep{Fowler} and the SSA \citep{WeisGreveHutter}.

\subsection{SSA version of the stress balance}\label{subsect:ssa}   When there is significant sliding at the base the driving stress is significantly balanced by membrane stresses \citep{HindmarshMembrane}.  ``Membrane stresses'', called ``longitudinal'' in flowline models, are deviatoric components $\tau_{11}, \tau_{22}, \tau_{12}$.   Any membrane stress balance is nonlocal because the driving stress is balanced by connection to distant ice either in the along-flow direction or at the margins of regions of fast flow.  The driving stress must be fully balanced by membrane stresses in cases where there is negligible traction at the base of the ice as in an ice shelf \citep{Morland}.

The simplest form of a membrane stress balance derivable from the Stokes model is the shallow shelf approximation \citep[``SSA''][]{WeisGreveHutter}.  It describes a fluid membrane of variable thickness in which gravity causes spreading.  When applied to ice streams the membrane also experiences traction (shear) forces on its base \citep{MacAyeal}.

We denote the velocity field which solves the SSA stress balance by $\bv=(v_1,v_2,v_3)$.  The horizontal velocity $(v_1,v_2)$ and the strain rates $D_{11},D_{22},D_{12}$ modeled by the SSA are independent of depth.  Denote vertically-averaged hardness by $\bar B$, so $\bar B\,H = \int_b^h B(T^*)\,dz$.  The vertically-averaged viscosity is
\begin{equation}\label{barnu}
  \bar\nu = \frac{\bar B}{2}\, \left[D_{11}^2 + D_{22}^2 + D_{11} D_{22} + \frac{1}{4} \left(D_{12} + D_{21}\right)^2\right]^{\frac{1-n}{2n}},
\end{equation}
\citep{MacAyeal}.  Following \citep{Morland,SchoofStream} we also define a vertically-integrated stress tensor
\begin{equation}\label{vertintstressSSA}
  T_{ij} = 2 \bar \nu\,H\, \left(D_{ij} + (D_{11} + D_{22})\delta_{ij}\right).
\end{equation}
for $i,j=1,2$.  Note $T_{11}+T_{22}$ is not generally zero so the membrane flow is conceptually ``compressible'' in the map-plane.  In fact $T_{11}+T_{22} = (2\bar \nu H)(-3 D_{33})$ so a negative value for $T_{11}+T_{22}$ is a map-plane ``pressure'' causing an upward strain rate $D_{33} = \partial v_3/\partial z >0$.

Let $(\tau_{b,1},\tau_{b,2})$ be the basal shear stress components applied to the base of the ice (modeled in next subsection).  The SSA is the pair of equations
\begin{equation}\label{deepSSA}
  \pp{T_{i1}}{x_1} + \pp{T_{i2}}{x_2} + \tau_{b,i} =  \rho g H \pp{h}{x_i}
\end{equation}
for $i=1,2$ \citep{SchoofStream}.  A slightly more concrete form is \citep{MacAyeal,WeisGreveHutter}
\begin{align}
 &\pp{}{x_1}\left[2\bar\nu H \left(2 \pp{v_1}{x_1} + \pp{v_2}{x_2}\right)\right]
 + \pp{}{x_2}\left[\bar\nu H \left(\pp{v_1}{x_2} + \pp{v_2}{x_1}\right)\right] \notag \\
 &+ \tau_{b,1} =  \rho g H \frac{\partial h}{\partial x_1}, \label{SSA} \\
 &\pp{}{x_1}\left[\bar\nu H \left(\pp{v_1}{x_2} + \pp{v_2}{x_1}\right)\right]
 + \pp{}{x_2}\left[2\bar\nu H \left(\pp{v_1}{x_1} + 2 \pp{v_2}{x_2}\right)\right] \notag \\
 &+ \tau_{b,2} = \rho g H \frac{\partial h}{\partial x_2}.\notag
\end{align}

Because our emphasis is on a novel interpretation of the SSA as a ``sliding law'', we note the heuristic which says Equation \eqref{SSA} is a splitting of the driving stress:
	$$\left(\begin{matrix} \text{membrane stresses} \\ \text{held by viscous} \\ \text{deformation}\end{matrix}\right) \, + \, \left(\begin{matrix} \text{stress held} \\ \text{at base by} \\ \text{till strength}\end{matrix}\right) = \left(\begin{matrix} \text{driving} \\ \text{stress}\end{matrix}\right).$$
In the extreme case of ice shelves the basal stress is zero so the left-most term fully supports the driving stress.  In the other (singular) extreme case of strong till none of the driving stress is held by the membrane strain rates, in which case the SSA predicts no flow \citep{SchoofStream}.  In that case we propose that the SIA should ``take over'' and predict flow by shear in horizontal planes.  Again, this is the sense of our title.

We will solve Equations \eqref{SSA} numerically at each time step to determine velocities $v_i$ from evolving geometry and temperature.  These equations are elliptic.  More precisely, these equations can be derived from a variational principle applied to a convex and bounded-below functional \cite[Equation (3.13) of][]{SchoofStream}, and the problem is well-posed.  On the other hand, we will add dynamics, that is we will couple the SSA with Equations \eqref{mapplanecont} and \eqref{conserve}.  Such an evolving coupled system is not yet known to be mathematically well-posed.

\subsection{Basal mechanics}\label{subsect:tillmech}  We follow \cite{SchoofStream} and assume that the basal shear stress has plastic form
\begin{equation}\label{plasticbase}
   \tau_{b,i} = -\tau_c\,\frac{v_i}{(v_1^2 + v_2^2)^{1/2}}
\end{equation}
for $i=1,2$, where $\tau_c$ is a positive scalar \emph{yield stress} \citep{Paterson}.  Conceptually, because the till is assumed to be plastic, it supports applied stress without deformation until the applied stress equals the yield stress, at which point some amount of deformation occurs.

In fact $\tau_{b,i}$ is only literally given by Equation \eqref{plasticbase} in map-plane locations where sliding is occurring.  In locations where no sliding occurs, the driving stress on the right side of \eqref{SSA} is equal to the left side in the sense that the basal stress equals the driving stress.  Where the base is not sliding, no division by zero occurs in the mathematically-precise form of Equations \eqref{SSA} because the problem is posed as minimizing a functional involving no division, though at its minimum the functional fails to have a well-defined gradient \cite[Equation (3.13)]{SchoofStream}.

In our thermomechanically-coupled model, the value of the till yield stress is partly determined by the availability of basal water.  The Mohr-Coulomb model for the yield stress of saturated till, which we adopt, is
\begin{equation}\label{MohrCoulomb}
   \tau_c = c_0 + (\tan\phi)\,(\rho g H - p_w).
\end{equation}
\cite[Chapter 8]{Paterson}.  Here $c_0$ is the till cohesion, $p_w$ is the pore water pressure (estimated below), and $\phi$ is a ``till friction angle,'' a strength parameter for the till comparable to ``angle of repose'' for granular piles.  In this paper we take $c_0=0$ partly for simplicity and partly because \cite{Paterson} notes an inconclusively broad observed range $0$--$40$ kPa for $c_0$.  The factor $\rho g H - p_w$ is the effective pressure of the overlying ice on the mineral portion of the saturated till.  The till is weakened by the presence of pressurized liquid water.

In our model $\tau_c$ necessarily represents the yield stress of the aggregate material at the base of an ice sheet, a poorly observed mixture of liquid water, ice, and granular till.

Recall we model an effective thickness $W$ of stored liquid water at the base of the ice column. This liquid water is in fact mixed with the solid part of the till, in some manner which we make no attempt to model.  Nonetheless we use $\Hmelt$ to estimate the pore water pressure:
\begin{equation}\label{porewatermodel}
   p_w = 0.95\, (\rho g H)\,\left(\frac{W}{W_0}\right).
\end{equation}
Note $0 \le W \le 2$ are imposed limits.  We use $W_0=2$ m throughout.  Thus we model the pore water pressure locally as at most a fixed fraction (95\%) of the ice overburden pressure $\rho g H$.  When the base is frozen we have $W=0$ so $p_w=0$ and the till is strong.  With a till friction angle $\phi = 15^\circ$ and fully-saturated till, the yield stress $\tau_c$ is 1.3\% of overburden, while with $\phi = 2^\circ$ the yield stress is 0.17\% of overburden.

\subsection{Combining the velocities}\label{subsect:super}  
\begin{figure}
\noindent\includegraphics[width=20pc]{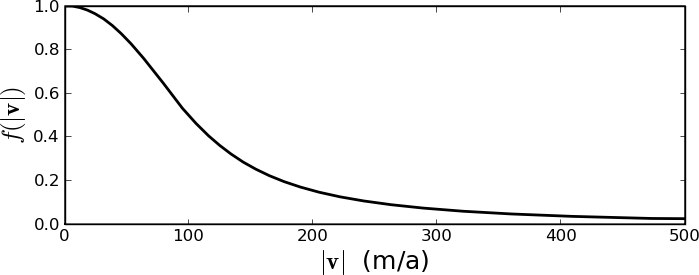}
\caption{Weighting $f(|\bv|)$ in Equation \eqref{superAverage}, versus sliding speed $|\bv|$.}
\label{fig:fofv}
\end{figure}

The use of an SSA result as the sliding velocity in an SIA model, the most novel aspect of our model, means combining the velocities computed by SIA and SSA to get the velocity used in mass continuity and conservation of energy.  Recall that $\bu=(u_1,u_2)$ is the SIA velocity (subsection \ref{subsect:sia}) and $\bv=(v_1,v_2)$ is the SSA velocity (subsection \ref{subsect:ssa}).  We compute the combined horizontal velocity $\bU=(U_1,U_2)$ by
\begin{equation}\label{superAverage}
(U_1,U_2) = f(|\bv|)\,\bu + (1-f(|\bv|))\bv,
\end{equation}
with $|\bv|^2 = v_1^2+v_2^2$ and
\begin{equation}\label{fofv}
f(|\bv|) = 1 - \frac{2}{\pi} \arctan\left(\frac{|\bv|^2}{100^2}\right).
\end{equation}
The weighting function $f$ has values between zero and one.  See Figure \ref{fig:fofv}.  It satisfies the general requirements of smoothness, monotone decrease, $f(|\bv|)\sim 1$ for $|\bv|$ small, and $f(|\bv|)\sim 0$ for $|\bv|$ large (e.g.~$|\bv|\gg 100$ m/a).  Many other choices satisfying these requirements are possible, and comparison of trusted Stokes model results to those from this paper is a potential method for choosing an optimal weighting $f(|\bv|)$.

It follows from Equations \eqref{superAverage} and \eqref{velocitySIA} that
\begin{equation}\label{combinedFlux}
\bQ = - f(|\bv|) D_{\text{SIA}} \grad h + (1-f(|\bv|)) \bv H,
\end{equation}
where
\begin{equation}\label{DSIA}
D_{\text{SIA}} = 2 (\rho g)^n\,|\grad h|^{n-1}\,\int_b^h A(T^*)\,(h-z)^{n+1}\,dz
\end{equation}
is the positive diffusivity associated to the thermomechanical SIA \citep[and references therein]{BBL}.  Note that the flux is a linear combination of a vector pointing down the surface slope and a vector determined through the SSA stress balance.  Only the former part of the flux is automatically diffusive for mass continuity Equation \eqref{mapplanecont}; compare \citep{Pattyn03}.  We address the numerical approximation of $\bQ$ and a time stepping scheme for Equation \eqref{mapplanecont} in subsection \ref{subsect:massconnumerics}.

Equation \eqref{SigmaFS} gives the strain heating rate $\Sigma_{\text{FS}}$ applicable to a full Stokes model.  We approximate $\Sigma_{\text{FS}}$ by a computable and appropriately shallow quantity $\Sigma$.  Strain rates for $\bu$ and $\bv$ do not sum in a simple way to yield the combined strain rates $D_{ij}$ for $\bU$, however.  Nonetheless we propose an approximate method for computing $\Sigma$ in pieces, as follows.  Denoting $D(\bu)_{ij} = (1/2) \left(\partial u_i/\partial x_j + \partial u_j/\partial x_i\right)$ and $D(\bv)_{ij} = (1/2) \left(\partial v_i/\partial x_j + \partial v_j/\partial x_i\right)$,
\begin{equation}\label{superForStrainRates}
D_{ij} \approx f(|\bv|) D(\bu)_{ij} + \left(1-f(|\bv|)\right) D(\bv)_{ij}.
\end{equation}
The SIA says that each of $D(\bu)_{11},D(\bu)_{12},D(\bu)_{21},D(\bu)_{22}$ are negligible $D(\bv)_{13},D(\bv)_{31},D(\bv)_{23},D(\bv)_{32}$ are all negligible in the SSA.  Thus equation \eqref{superForStrainRates} can be rewritten
\begin{align*}
&\left(D_{ij}\right) \approx f(|\bv|) 
\begin{pmatrix} 0 & 0 & D(\bu)_{13} \\
 & 0 & D(\bu)_{23} \\
 &  & 0
\end{pmatrix} \\
&\quad+ \left(1-f(|\bv|)\right) \begin{pmatrix} D(\bv)_{11} & D(\bv)_{12} & 0 \\
 & D(\bv)_{22} & 0 \\
 &   & D(\bv)_{33}
\end{pmatrix}.
\end{align*}
Only entries on and above the diagonal need be displayed because of symmetry.  It follows from \eqref{superForStrainRates} that
\begin{equation*}
D^2 = f(|\bv|)^2 \left(D(\bu)_{13}^2 + D(\bu)_{23}^2\right) + \left(1-f(|\bv|)\right)^2 D(\bv)^2
\end{equation*}
where $D(\bv)^2 = D(\bv)_{11}^2 + D(\bv)_{22}^2 + D(\bv)_{12}^2 + D(\bv)_{11} D(\bv)_{22}$.  This yields a computable approximation of $\Sigma_{\text{FS}}=2 B(T^*) D^{(1+n)/n}$, in terms of quantities available in the SIA and SSA stress balance computations.

\section{Numerics and verification}\label{sect:numerics}

Each part of the continuum model of section \ref{sect:model} is numerically approximated in PISM, an open source project (\texttt{www.pism-docs.org}).  For the time-dependent, thermo-mechanically-coupled, non-sliding SIA, the numerical schemes used in PISM are described in \citep[Appendix A]{BBL}.  The PISM schemes described in this section are those additional schemes which were used to produce results for this paper.

All PISM numerical schemes share these properties:\renewcommand{\labelenumi}{(\emph{\roman{enumi}})} \begin{enumerate}
\item they use finite difference approximations on a regular, rectangular grid in horizontal variables $x_1,x_2$,
\item they do not use a rescaling of the vertical axis \citep[e.g.~as in][]{Jenssen}, but instead model the ice surface as a boundary between ice and air within the computational grid,
\item time stepping is adaptive because the stability conditions appropriate to each explicit or semi-implicit time-stepping numerical scheme are combined to give a next time-step based on the current geometry, velocity, and temperature fields, and
\item the PETSc library \citep[Portable, Extensible Toolkit for Scientific computation; ][]{petsc-user-ref} manages parallel communication and the parallel, iterative solution of linear systems.\end{enumerate}

\subsection{Approximation of the SSA}\label{subsect:numssa}  Equations \eqref{SSA} are nonlinear in the velocities $v_i$.  Therefore we solve them through an iteration in which the values of the velocities from the previous time step are used to compute an initial estimate of the vertically-integrated effective viscosity $\bar \nu^{[0]}$.  The iteration then solves the linearized form of Equation \eqref{SSA} to compute new velocities and updated viscosity $\bar \nu^{[k]}$ with each iteration.  The final values $\bv$ from this iteration are the ``SSA velocities''.  Such an iteration has been widely used to solve the diagnostic problem of computing the flow velocity in ice shelves \citep{MacAyealetal}.  Appendix A details the PISM parallel implementation.

We regularize Equations \eqref{barnu} and \eqref{plasticbase} to avoid divisions by zero.  In fact, we must regularize the problem to even think of it as a system of partial differential equations, because otherwise the correct formulation of the problem is a weaker form (a ``variational inequality'').  Following \citep{SchoofStream} we introduce both an ice viscosity regularization and a till plasticity regularization.  For the former, we replace Equation \eqref{barnu} by
\begin{equation}\label{barnureg}
  \bar\nu^\eps = \frac{\bar B}{2}\, \left[\frac{\eps^2}{L_\nu^2} + D_{11}^2 + D_{22}^2 + D_{11} D_{22} + \frac{1}{4} \left(D_{12} + D_{21}\right)^2\right]^{\frac{1-n}{2n}}
\end{equation}
where $\eps$ has units of velocity and $L_\nu$ is a characteristic horizontal length scale for the membrane stresses.  Our values $\eps = 1$ meter per year and $L_\nu=1000$ km give $\eps/L_\nu \approx 3 \times 10^{-14}\,\text{s}^{-1}$, a strain rate which should be compared to the typical membrane strain rates in a ice stream or shelf.  These typical rates are of order $100$ meters per year per $100$ km, about $3 \times 10^{-11} \,\text{s}^{-1}$, in the Siple Coast ice streams \citep{MacAyeal} or the Ross ice shelf \citep{MacAyealetal}.  Because of the squares in Equation \eqref{barnureg}, for large parts of the sliding domain there is a $10^6$ difference in magnitudes between the regularizing constant $(\eps/L_\nu)^2$ and the largest of the other quantities in square brackets.

For the plastic regularization we modify Equation \eqref{plasticbase} by choosing a small velocity $\delta = 0.01$ meters per year and defining
\begin{equation}\label{plasticbasereg}
   \tau_{b,i}^\delta = -\tau_c\,\frac{v_i}{(\delta^2 + v_1^2 + v_2^2)^{1/2}}.
\end{equation}
This regularization is only significant at or near locations where there is no sliding in the perfectly plastic model.  Described another way, we have made the till linearly viscous $\tau_{b,i} = -\beta v_i$, with a very large coefficient $\beta = \tau_c \delta^{-1}$, when sliding velocities are of order $\delta$ or smaller.  Thus the ice experiences slight sliding even where the till would not fail in the perfectly plastic limit.  It follows from Equation \eqref{MohrCoulomb} that this sliding is of order $\delta$ where the base is melted but not yielding, and about $10^{-2}\delta$ where the base is frozen and not yielding.

In the context of a higher-order \cite{Blatter} model, \cite{SchoofCoulombBlatter} has recently shown that the ice flow velocity which results from the plasticity-regularized equations converges to the purely-plastic result.  We expect that a similar result applies in our vertically-integrated, thermomechanically-coupled, and geometry-evolving case.

\subsection{Verification of the SSA numerics}\label{subsect:verifssa}  We use an existing exact solution for primary verification of our numerical approximation of the ``diagnostic'' isothermal SSA (including plastic till but without time evolution of stream geometry).  ``Verification'' is used here as in computational fluid dynamics \citep{Wesseling} to mean the comparison of numerical results to exact predictions of the continuum model.

\begin{figure}
\noindent\includegraphics[width=20pc]{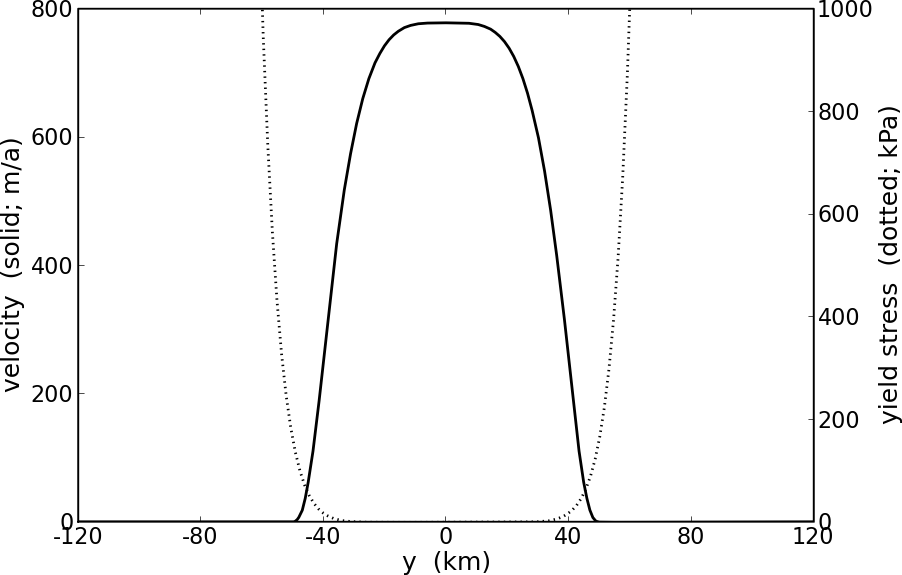}
\caption{Solid curve: Ice stream velocity $v(y)$in the down slope direction in the exact SSA solution.  Dotted: Till yield stress $\tau_c(y)$ grows sharply at $|y|=40$ km, and so significant sliding occurs only within the interval $-40\,\text{km} < y < 40\,\text{km}$.  The region of sliding is not predetermined.}
\label{fig:velexactI}
\end{figure}

The exact solution we use appears in section 4 of \citep{SchoofStream}, but it essentially arises in the analysis in \citep{Raymondenergy} as well.  Parameters and notation for the exact solution are in Table \ref{tab:testIconst}.  Because of translational symmetry in the downstream direction, the SSA with plastic till reduces to a one variable problem in the cross-stream coordinate $y$.  The solution is a formula $v(y)$ for the velocity which solves a free boundary problem describing a single ice stream with given till yield stress values $\tau_x(y)$.  Because this is a free boundary problem, the width of the sliding ice stream is found simultaneously with the velocity profile $v(y)$.  Note that the thickness, bed slope, and surface slope are all constant.  We use the particular instance illustrated in Figure \ref{fig:velexactI}; it is called ``Test I'' in PISM \citep{pism-user-manual}.   

\begin{figure}
\noindent\includegraphics[width=20pc]{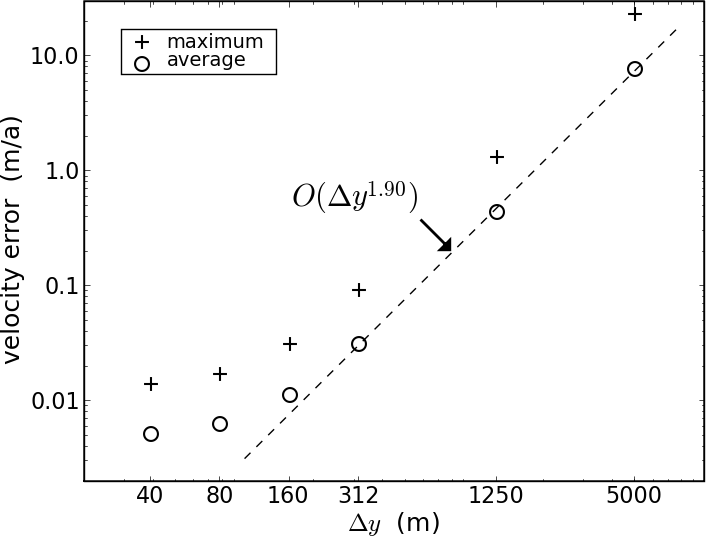}
\caption{Numerical errors in along-flow velocity for exact SSA solution shown in Figure \ref{fig:velexactI}.  Convergence at rate $O(\Delta y^{1.90})$ is found by fitting errors for grids with $160 \le \Delta y \le 5000$ m.}
\label{fig:velerrsI}
\end{figure}

Our goal is to measure how well the finite difference scheme, described in Appendix A, solves this free boundary problem.  Using a refinement path with $\Delta y$ decreasing from $5$ km to $40$ m, the maximum and average numerical errors in velocity decay as shown in Figure \ref{fig:velerrsI}.  The numerical error at grid point $k$ is $|v_{\text{num}}(y_k)- v_{\text{exact}}(y_k)|$.  The maximum of these pointwise errors converges rapidly down to a level of less than $10^{-1}$ m/a as the grid is refined.  For comparison, the exact maximum velocity is $777.5$ m/a.  The average error is about three times smaller than the maximum error.  As shown in Figure \ref{fig:velerrsI}, for ``rough'' grids with $160\,\text{m} \le \Delta y \le 5\,\text{km}$ the errors decay at almost the optimal $O(\Delta y^2)$ rate for such finite difference schemes \citep{MortonMayers}.  For the finest two grids in Figure \ref{fig:velerrsI} the errors are significantly influenced by the viscosity iteration relative tolerance.  This tolerance has been set to $5 \times 10^{-7}$ and the linear iteration (Krylov solver) relative tolerance to $10^{-12}$ for these verification experiments.

With regard to this SSA free boundary problem, \cite{SchoofStream} also demonstrates that for any aspect ratio that could conceivably be regarded as ``shallow'', such as depth-to-width ratios of 0.1 or smaller, the velocity predictions of the SSA are very close to those of the full Stokes equations (applied to the same boundary values).

\subsection{Approximation of mass continuity}\label{subsect:massconnumerics}  The mass continuity Equation \eqref{mapplanecont} is only strictly diffusive if the vertically-integrated flux $\bQ$ is exactly in the downslope direction at all times and all locations on the ice sheet.  But there is no reason to expect this to be true for real ice sheets.  For instance, in ice streams with minimal bed topography and low surface slope it is likely that the flow direction is controlled largely by side drag and other effects unrelated to local surface slope.  In our model some flux acts diffusively and some is treated as generic transport.

In fact, recalling Equation \eqref{combinedFlux}, it follows that Equation \eqref{mapplanecont} can be written
\begin{equation}\label{massconform}
\ddt{H} = \Div\left(\tilde D\grad h\right) - \tilde\bv \cdot \grad H - (\Div\tilde\bv) H + (M - S).
\end{equation}
where $\tilde D = f(|\bv|) D_{\text{SIA}}$ is a positive scalar and $\tilde\bv=(1-f(|\bv|)) \bv$.  Thus, temporarily ignoring the distinction between $H$ and $h$, Equation \eqref{massconform} has diffusion, advection, reaction, and source terms on the right side (in that order).

Our numerical approach to Equation \eqref{massconform} is to use the simplest explicit finite difference method which is conditionally stable and consistent \citep{MortonMayers}.  Though we describe the scheme next as though there were only one horizontal variable $x$, recovery of the $(x_1,x_2)$ form is straightforward.  We will use $m$ in the superscript to denote time step $t^m$ and $j$ in the subscript to denote horizontal grid point $x_j$.  Also suppressing the superscript $m$ on the right because all terms are evaluted at $t^m$ (i.e.~explicitly), the scheme for \eqref{massconform} is
\newcommand{\Up}[2]{\text{Up}\left({#1}_\bullet\big|{#2}\right)}
\begin{align}
&\frac{H_j^{m+1} - H_j^m}{\Delta t} = \frac{\tilde D_{j+1/2} \left(h_{j+1}-h_j\right) - \tilde D_{j-1/2} \left(h_j-h_{j-1}\right)}{\Delta x^2} \label{massconformFD} \\
&\qquad - \tilde v_j \frac{\Up{H}{\tilde v_j}}{\Delta x}- \left(\frac{\tilde v_{j+1} - \tilde v_{j-1}}{2\Delta x}\right) H_j + (M_j - S_j). \notag
\end{align}
The weighted diffusivity $\tilde D_{j+1/2}$ is computed by the \cite{Mahaffy} scheme as in \citep{BBL}.  Our upwind scheme uses notation $\Up{H}{\tilde v_j}$ for $\tilde v_j(H_{j}-H_{j-1})$ if $\tilde v_j \ge 0$ and $\tilde v_j(H_{j+1}-H_{j})$ if $\tilde v_j < 0$.

Scheme \eqref{massconformFD} is conditionally stable.  A condition applies to the explicit diffusion scheme: $(\max_j \tilde D_{j+1/2}^m)\, \Delta t < 0.5 \Delta x^2$ \citep[chapter 2]{MortonMayers}.  The CFL condition for the upwind scheme applies: $(\max_j \tilde v_j^m)\,\Delta t < \Delta x$ \citep[chapter 4]{MortonMayers}.  We adaptively take $\Delta t$ to be the largest time step which satisfies both of these conditions at every point on the grid.  Though the mass continuity equation is strongly nonlinear, stability follows because the maximum principle applies \citep[pages 16--18, 36--38, 50--51, and 94--96]{MortonMayers}.

There is an additional condition for the stability of the conservation of energy scheme \citep{BBL}, and it is also included in the adaptive time-stepping scheme.

\section{Experiments}\label{sect:setup}  

The model described above is not applied to a real ice sheet in this paper, but it is being applied to the whole Antarctic ice sheet in separate work (in preparation).  Here we examine how the modeled flow speed and ice sheet volume depend on certain parameters.  We use four experiments which explore part of the parameter space and we study grid refinement.  These experiments demonstrate the stability of the model with respect to parameter changes.

Each of the four experiments starts from the same circular ice sheet which is the initial state.  It is very similar to the steady state (i.e.~at 200 ka model years) of experiment A in EISMINT II \citep{EISMINT00}, which is denoted ``EISIIA''.  The extent of the model domain and the net surface mass balance and  surface temperature maps are all exactly as in EISIIA, and our initial state comes from a model run using only the non-sliding, thermomechanically-coupled SIA, as in EISIIA.  Two differences are that ours includes a bedrock thermal model (Equation \eqref{conserveLitho}) and modeled basal water effective thickness (Equation \eqref{Hmelt}).  For our initial state, the basal melt rate is computed by Equation \eqref{basalmeltrate}, and it \emph{is} included in the mass continuity Equation \eqref{mapplanecont} and in the vertical velocity (Equation \eqref{vertvel}).  Our initial state is on a 10 km horizontal grid, versus 25 km for EISIIA.

The initial state and the experiments which follow use an unequally-spaced vertical grid in the ice (and air above, up to elevation 5000 m), with spacing increasing from 12.9 m at the bed to 87.1 m at the top.  The vertical grid in bedrock has forty equal 12.9 m grid spaces.

\begin{figure}
\noindent\mbox{\parbox{39pc}{\includegraphics[width=19pc]{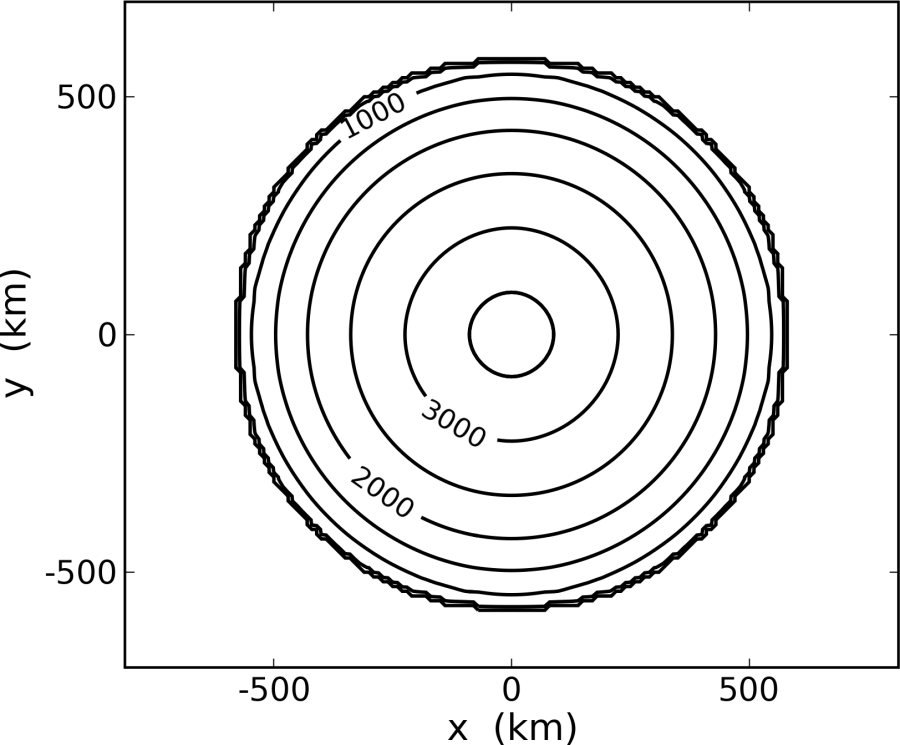} \quad \includegraphics[width=16pc]{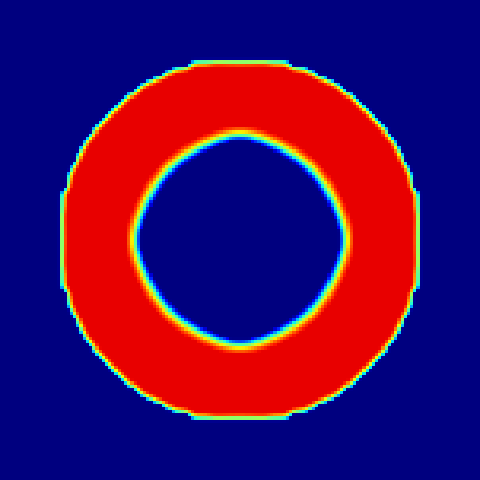}}}
\caption{Ice thickness (left) and thickness (right) of basal water layer for the initial state.  Effective basal water thickness $W$ ranges from zero (blue) to 2 m (red); the central blue has frozen base and zero basal water.  The right figure, and all other color map-plane Figures in this section, show a $1500 \times 1500$ km square region.}
\label{fig:P0A}
\end{figure}

The numerically-computed initial state has radius about 575 km, center thickness $3708.75$ m, center (absolute) temperature at the ice base 256.24 K, and volume $2.208\times 10^6\,\text{km}^3$.  Figure \ref{fig:P0A} shows the thickness and the extent of melted base for the initial state.

This initial ice sheet does not come from a sliding model, but it is ``ready to slide'' in locations where the till yield stress $\tau_c$ is sufficiently low.  On the one hand, if the till friction angle were set to a large value everywhere (e.g.~$15^\circ$ or greater), then our model reduces to the non-sliding thermomechanically-coupled SIA with no sliding.  On the other hand, if the till friction angle were set everywhere to a sufficiently weak value (e.g.~$5^\circ$) then the sheet would slide in all radial directions.  In fact, however, each of our four experiments includes a map of till friction angle $\phi$ with weak strips.  Under most of the sheet $\phi=15^\circ$, while in the weak strips we set $\phi=5^\circ$, except in experiment P4 which explores this weak till parameter.  The experiments start from the initial sheet described in the previous paragraphs, and run for 5000 model years.  One run, experiment P1 on a 10 km grid, is extended to 100k model years.  See Table \ref{tab:experiments}.

The spatial distribution of weak till for three of the four experiments is shown in Figure \ref{fig:tillmaps}.  Experiment P1 has four weak till strips with widths $30$, $50$, $70$, and $100$ km, oriented in the cardinal directions.   The cross stream profiles of till friction angle $\phi$ for the four strips are shown in Figure \ref{fig:crossprofileP1phi}.  Experiment P2 has three weak till strips each with width $70$ km, but oriented at $0^\circ$, $10^\circ$, and $45^\circ$ relative to the closest cardinal grid direction.  The goal of this experiment is to determine to what degree the grid alignment of the weak strip affects model outcome.  Experiments P3 and P4 each have four weak strips in the cardinal directions with identical widths $70$ km.  Experiment P3 has constant weakness of $5^\circ$, with cross-stream profile exactly as shown in the $70$ km width case in Figure \ref{fig:crossprofileP1phi}, but there is differing bed slope in the four streams.  We essentially use the ``trough'' bed topography from EISMINT II experiment I \citep{EISIIdescribe}, but with total elevation drops of $0$, $500$, $1000$, and $2000$ m in the 650 km length of the troughs, giving slopes of $0$, $0.077$, $0.154$, and $0.308$ percent, respectively.  Figure \ref{fig:P3bed} shows the map of bed elevation for P3.  Finally, experiment P4 has different values for the downstream value of $\phi$.  ``Downstream'' starts 400 km from the beginning of the weak strip.  The different strips have downstream $\phi$ values $2^\circ$, $3^\circ$, $8^\circ$, and $10^\circ$, respectively, while the upstream value is $5^\circ$ for all strips.

All experiments are grounded ice sheets without calving fronts.  The combined SIA-SSA model in this paper is applied at all points and there is no calving-front boundary condition.  We have not analyzed the asymptotic shape of the resulting grounded margin for the combined model.  The  margin shape arising from the isothermal and thermomechanically-coupled nonsliding SIA models is analyzed in \citep{BLKCB,BBL}, while numerical scheme improvements for SIA margins appear in \citep{SaitoMargin}, but these results do not apply directly to our case.

In preparation for the next section, and to illustrate why the initial state is ``ready to slide'' in the weak till strips, in Figure \ref{fig:stresscompareP1start} we compare the magnitude of the basal driving stress $\rho g H |\grad h|$, for the initial state, to the till yield stress computed from Equation \eqref{MohrCoulomb} for the till friction angle map of experiment P1.  We see that the driving stress exceeds the yield stress only in the portions of the strips where the base has stored melt water; compare Figure \ref{fig:P0A}.  The frozen parts of the weak till strips, though they have till friction angle $\phi=5^\circ$, actually have large yield stresses as a consequence of Equation \eqref{MohrCoulomb}.  In any case, the regions where the driving stress exceeds the yield stress are expected to slide, although the sliding velocity is controlled by membrane stress connections to nonsliding parts.  Because the difference between driving and yield stress is large within the downstream portions of the weak strips, the sliding will initially be very rapid.

\section{Results}\label{sect:results}

The results in this paper can be reproduced by running a script included in the PISM source code distribution.  The combined time of all of our results is roughly 4600 processor-hours on a mix of 2.3 GHz quad core Xeon and 2.6 GHz dual core AMD Opteron processors.  Runs used 128 processors simultaneously for the 5 km grid results.

\subsubsection*{Parameter dependence}  A basic way to illustrate  and compare flow is to show the surface velocity at a fixed time as in Figure \ref{fig:PallSpeed}.  In all cases the portions of the weak till strips corresponding to significant basal melt produced significant sliding (e.g.~$\ge 100$ m/a) at some point in the 5000 year runs, but this snapshot at the final time shows some streams are ``off''.  Note that in all runs the peak surface speed outside of the weak strips, namely the surface speed from the SIA, is approximately 60 m/a.

The time evolution of the modeled flow must be illustrated in a different way.  A highly-averaged measure is the volume of the whole ice sheet.  As shown in Figure \ref{fig:PallVolTS} the volume decreases in every experiment.  The rate of volume loss is most rapid in the first 500 model years because the high driving stresses in the initial state produce fast sliding into the ablation area.  This fastest initial flow is self-limiting because the geometry changes to produce lower driving stresses, and because cold ice is advected down into the stream regions, with the ultimate effect that the modeled basal melt rate decreases or even becomes negative (refreeze), so that the modeled till yield stress eventually increases at some future time in each stream.

Experiment P1 was continued to a 100 ka run at 10 km resolution.  The volume time series shown in Figure \ref{fig:P1VolTScont}.  The ``leveling out'' suggested in the 5 ka result is realized and an approximate steady state is achieved.  There is sliding flow at speeds typical of real ice streams at every time during this longer run, as shown in Figures \ref{fig:P1DownSpeedcont} and \ref{fig:P1DownSpeedcontWidth100}.  Those Figures show the spatial average of the vertically-integrated horizontal velocity over the positive thickness, downstream portion of each weak strip.  There is rapid flow at speeds in excess of $10^4$ m/a in the first few model years, but these high initial speeds are caused by the high driving stresses created by the non-sliding run which created the initial state.  The modeled flow appropriately re-stabilizes to reasonable sliding speeds.  The 100 km wide stream in experiment P1 produces the most volatile flow.  The explanation for this is that the center of the 100 km strip periodically cools and refreezes, causing shutdown of the central part of the sliding flow.  Oscillations appear in the area-averaged flow speed, with period about 900 a (inset in Figure \ref{fig:P1DownSpeedcontWidth100}).

The Figures showing flow speeds all include large and very brief excursions to much faster flow; these are vertical lines sometimes going off scale.  These correspond to discrete advance and retreat of the margin by a single or a few grid spaces.  We believe that such discrete margin movement results in a brief reduction of backpressure, followed by fast flow and advance of the margin to again ``press'' against the few essentially non-sliding ice-filled grid spaces at the margin and in the ablation area.

We now show a sequence of Figures (\ref{fig:P1DownSpeed},\ref{fig:P2DownSpeed},\ref{fig:P3DownSpeed},\ref{fig:P4DownSpeed},\ref{fig:P4bwat},\ref{fig:P4at5yr_csurf}) from the highest resolution (5 km) runs.  

Figure \ref{fig:P1DownSpeed} shows that the two narrower width streams in experiment P1 reach a level of relatively steady flow at average speeds of roughly 100 m/a within the first 1 ka.  By contrast, the two wider streams enter cycles with average speeds varying from under 10 to over 100 m/a.

Recall that experiment P2 examines how the flow depends on the grid orientation of weak strips, relative to the cartesian grid directions.  As shown in Figure \ref{fig:P2DownSpeed}, although the strength and width of the weak strips are the same, the resulting speeds are somewhat different.  There is rough agreement for the first 500 model years.  The range of speed variation (20 to 200 m/a) roughly agrees for the whole 5000 years.  All three streams enter similar oscillating states, but the cycles are out of phase and differ in period (1300 to 2100 years).  Note that because all three streams are modeled within the same ice sheet there is some ``competition'' for ice near the divide (dome).  Nonetheless be believe grid orientation has some effect on flow speed, a result which needs to be clarified in future work.

Experiment P3 evaluated the effect of bed slope on the modeled flow.  We see in Figure \ref{fig:P3DownSpeed} that larger slopes generate faster flow speeds and greater variability.  The period of the cycle is only weakly dependent on bed slope, however.

Experiment P4 evaluated the effect of till friction angle in the downstream region of the till.  Figure \ref{fig:P4DownSpeed} shows that smaller till friction angle yields greater volatility in speed, with higher highs and lower lows.  The variability of the modeled stream flow certainly relates to the availability of basal water, which modulates till yield stress (subsection \ref{subsect:tillmech}).  Figure \ref{fig:P4bwat} shows the basal water thickness $\Hmelt$ at the end of the 5 ka run for P4.  It should be compared to the P4 surface speed shown in Figure \ref{fig:PallSpeed} which shows only one active ice stream.  The only active stream is the one which is underlain by the maximal amount of basal water ($\Hmelt = 2$ m), which is therefore the stream with the lowest till yield stress.  Though the yield stress is lowest in the active stream, the till friction angle is slightly large in the active stream ($3^\circ$) than in the stream with the lowest friction angle ($2^\circ$).

Like Figure \ref{fig:PallSpeed}, Figure \ref{fig:P4at5yr_csurf} also shows surface speed for experiment P4, but very near the beginning of the run (at 5 a).  We see rapid sliding as the sheet reduces its driving stress to match the (recently and suddenly) reduced basal resistance which occurs at the start of all our experiements.  In fact we see that the fastest flow at this time is in excess of 3000 m/a, and the weakest till corresponds directly to the lowest till friction angle because basal water thickness has not yet lost its initial angular symmetry (e.g.~as in Figure \ref{fig:P0A}).

The surface elevation evolves though the mass continuity equation in all runs.  Because the sliding is rapid, surface evolution occurs rapidly and is mostly stabilized within the first 1 ka.  Figure \ref{fig:P1vsP0A} illustrates the fact that the high driving stresses of the initial state have evolved, by the end of the 100 ka run, to lower driving stresses.  There is roughly constant surface slope along the sliding portion of the stream; compare Figure 2 in \citep{Joughinetal2001} which shows the surface profile of along approximate flowlines in the NE Greenland ice stream.  

\subsubsection*{Effects of grid refinement}  Figure \ref{fig:gridP1vol} shows that experiment P1 behaves stably under changes in grid spacing when measured by volume.  Over the full 5000 year evolution there is no obvious convergence toward a fine grid result, but the evidence for convergence is clear when we consider only the first 1000 years of the run, as in Figure \ref{fig:gridP1vol1ka}.  In particular, the 7.5 and 5 km grid results remain very close for the first 500 years of the run.  Grid refinement results for experiments P2, P3, and P4 are very similar and are not shown.

The vertical grid was described in the previous section and was used in all runs except for one version (10 km horizontal grid) wherein the number of grid points in the vertical was increased by a factor of two.  This is labeled ``fine vert'' in the last two Figures.  This change in the vertical grid makes little difference in these Figures and to essentially all results in this paper.

Grid refinement produces good behavior in terms of the spatial distribution of surface speed in modeled ice streams.  Figure \ref{fig:P1speeddetail} shows the final state of the 70 km wide stream in experiment P1, at four horizontal grid spacings.

\subsubsection*{Parallelization}  In a parallel application like PISM, ideally a doubling of the number of processors would halve the time to complete a run.  This is never quite achieved in practice for any parallel application because of the time for interprocess communication and other ``overhead''.  In our case the parallel solution of linear systems is the most difficult part to speed up through parallelization, but this is handled in a sophisticated manner by the PETSc library \citep{petsc-user-ref}.  As shown in Figure \ref{fig:timingP1parallel}, a perfect rate of speedup is not achieved but parallelization is notably effective anyway.  The change to  ``asymptotically almost perfect'' appearance of the speedup in the range from 16 to 128 processors is likely an effect of memory parallelism, wherein the part of the simulation handled by a processor becomes smaller as the number of processors increases, so that it fits in fast cache memory.

\section{Discussion}\label{sect:discussion}

The experiments above are for a simplified ice sheet with a steady and angularly-symmetric climate (surface mass balance and temperature).  On the other hand, new coupling mechanisms interact in a highly dynamic manner, as appropriate to real ice sheet dynamics, so results are not easily summarized.  Nonetheless we can list some common features of the above results:
\begin{enumerate}
\item All experiments start with a brief period of very fast sliding.  In the initial state the driving stress is held fully by resistance at the base, but once sliding is allowed the driving stress is far out of balance.  As shown in Figure \ref{fig:P4at5yr_csurf}, for example, the surface velocity exceeds 3000 m/a at 5 years into experiment P4, but it later stabilizes to speeds typical of large ice streams with low bed slope (e.g.~Figure \ref{fig:PallSpeed}).
\item After an initial period of rapid ice volume loss, caused by fast sliding into the ablation area, the ice sheet volume stabilizes.  Fitting an exponential decay curve to the data shown in Figure \ref{fig:P1VolTScont} gives 22 ka as the exponential time for volume decay.  The estimated steady volume is $2.106\times 10^6$ $\text{km}^3$ in experiment P1, very slightly smaller than the mean steady volume of $2.128\times 10^6$ $\text{km}^3$ in EISMINT II experiment A \citep{EISMINT00}.
\item Surface velocity within ice streams sometimes enters a limit cycle.  Wider ice streams are more likely to continue cycling because the center of the stream is the location at which advected cold ice can cause basal refeeze (followed by an increase in till yield strength and sliding shut down).  In a wider stream there is more space for advection to dominate the energy balance, away from the intense dissipation heating at stream margins.
\item Experiments without a bedrock thermal layer (not shown) suggest that having a bedrock layer of modest thickness (e.g. $\ge 100$ m) is important.  Without it the basal lubrication dynamics are more erratic.  The need for a modest layer identified here, to mollify short timescale variations in basal lubrication, is distinct from the known importance of a relatively deep bedrock thermal layer in order to correctly assimilate long time scale climate variations into the ice sheet temperature field.
\item \cite{SchoofStream} observes that for the plastic till version of the SSA, the stability of time-dependent geometry evolution is unknown as a mathematical matter.  In fact there is no proof that cliffs will not appear on the ice surface at sliding/nonsliding transition locations in the \cite{SchoofStream} model, if it were to be used by itself in a time-dependent manner.  Our model adds horizontal plane shearing by the SIA, which causes the equation of mass continuity to be partly diffusive, which smooths the surface.  It therefore exhibits stable geometry evolution.
\end{enumerate}

The fundamental explanation for the variability of the stream flow in our model is exactly the mechanism described in \citep{Raymondenergy}, in which strain dissipation and sliding friction heating compete with a combination of cold ice advection and basal drainage, in providing lubrication for sliding flow.  In our case, basal drainage is modeled only through limiting the effective basal water thickness to 2 m.

We note three caveats about our model, beyond the shallowness assumptions addressed in the introduction:
\renewcommand{\labelenumi}{(\emph{\alph{enumi}})}
\begin{enumerate}
\item Experiments in this paper do not include floating ice shelves, which are subject to zero till yield stress and to the floatation criterion.  On the other hand, the same PISM code for solving the SSA has successfully duplicated the EISMINT diagnostic comparison of modeled to observed Ross ice shelf velocities \citep{pism-user-manual,MacAyeal}.
\item Modeling the flow of ice in the immediate vicinity of the grounded margins apparently requires more complete stress balances than used here.
\item Detailed modeling of ice stream shear margins is likely to require more complete stress balances than used here.  Our model allows transmission of side resistance into the interior of the sliding ice stream through membrane stresses only.  Real shear margins involve bridging stresses, exotic concentration of strain dissipation heating and crevasse-related cooling \citep{TrufferEchelmeyer}, and accumulation of anisotropy and damage in the ice fabric, all missing in our model.
\end{enumerate}
These caveats each suggest the importance of high resolution, thermomechanically coupled ``higher-order'' and (full) Stokes stress balance modeling, either at whole ice sheet scale or in locally-refined submodels within a shallower whole sheet model like the one here.

\section{Conclusion}\label{sect:conclusion}

The open source model in this paper includes membrane stress balance, is thermo-mechanically coupled, includes a basal strength parameterization depending on basal melt, and is high resolution.  We demonstrate runs for a Greenland scale ice sheet on 5 km grid for a duration of 5000 model years, and for 100 ka on a 10 km grid, so a model with these features is now available for most prognostic modeling purposes.

In fact, our model is of intermediate computational expense between the easier thermomechanically-coupled SIA and more expensive ``higher-order'' models.  Its implementation is fully parallel.  Specifically, we believe that the numerical and scientific computing choices made here are an effective step toward parallel implementation at high resolution of the \cite{Blatter} model with a treatment of basal sliding following \citep{SchoofCoulombBlatter}.

Our model has distinct advantages over most existing whole ice sheet models which include sliding.  The basal sliding velocity field here is continuous, unlike the sliding field generated by the SIA with a classical temperature-dependent sliding law.

The underlying physical model is supported by recent observations \citep{JoughinMacAyealTulaczyk,Tulaczyketal2000} and theory \citep{SchoofStream} for ice streams.  The resulting flow speeds are realistic for those parts of real ice sheets where bed gradients are modest, including the NE Greenland and Siple coast ice streams.  Our model is appropriate to the shallow southern margins (lobes) of the Laurentide ice sheet, which experienced significant streaming.

We expect that this model can assimilate the information in basal shear stress maps derived by inverse modelling, though the precise method for deriving till friction angle from inversion-derived basal shear stress and velocity remains to be addressed.  The heuristic concept that till is weaker at locations of lower bed elevation, because of a marine history for that bed \citep{HuybrechtsdeWolde}, can be directly implemented, however.  This yields a promising model of West Antarctic ice streams within a whole Antarctic ice sheet model (in preparation).  Short time-scale, local changes basal lubrication of the type associated to moulin drainage of surface melt \citep{Zwallyetal02} can also be simulated directly without inverse methods, by forcing changes in modeled pore water pressure.

\appendix

\section{Finite differences for SSA stress balance}  We use the notation of subsection \eqref{subsect:ssa}, but let $x=x_1$ and $y=x_2$ and we denote an approximation of the function value $f(x,y)$ at a grid point $(x^i,y^j)$ by $f^{i,j}$.  We use difference notation $\delta_{+x}f^{i,j} = f^{i+1,j}-f^{i,j}$, $\delta_{-x}f^{i,j} = f^{i,j}-f^{i-1,j}$, and $\Delta_{x}f^{i,j} = f^{i+1,j}-f^{i-1,j}$, and corresponding notation for $y$ differences.  Temporarily denoting the product $\bar\nu\,H$ by ``$N$'', our approximation of the first of Equations \eqref{SSA} is
\begin{align}
&2 \frac{N^{i+\frac{1}{2},j}}{\Delta x} \left[2\frac{\delta_{+x}v_1^{i,j}}{\Delta x} + \frac{\Delta_{y} v_2^{i+1,j} + \Delta_{y} v_2^{i,j}}{4 \Delta y}\right]\label{fdssaeqn1} \\
&\quad\quad - 2 \frac{N^{i-\frac{1}{2},j}}{\Delta x} \left[2\frac{\delta_{-x}v_1^{i,j}}{\Delta x} + \frac{\Delta_y v_2^{i,j} + \Delta_y v_2^{i-1,j}}{4 \Delta y}\right] \notag \\
&+ \frac{N^{i,j+\frac{1}{2}}}{\Delta y} \left[\frac{\delta_{+y} v_1^{i,j}}{\Delta y} + \frac{\Delta_x v_2^{i,j+1} + \Delta_x v_2^{i,j}}{4 \Delta x}\right] \notag \\
&\quad\quad - \frac{N^{i,j-\frac{1}{2}}}{\Delta y} \left[\frac{\delta_{-y}v_1^{i,j}}{\Delta y} + \frac{\Delta_x v_2^{i,j} + \Delta_x v_2^{i,j-1}}{4 \Delta x}\right] \notag \\
&+ \tau_{b,1}^{i,j} = \rho g H^{i,j} \frac{\Delta_x h^{i,j}}{2\Delta x}. \notag
\end{align}
The second of Equations \eqref{SSA} is approximated by a similar scheme.  In computing the staggered value of $\bar\nu\,H$ the thickness $H$ is averaged onto the staggered grid, while staggered grid values of the effective viscosity $\bar\nu$ come from Equation \eqref{barnu} using the same centered finite differences as in Equation \eqref{fdssaeqn1}.  Figure \ref{fig:ssastencil} shows which of the quantities $\bar\nu\,H,v_1,v_2$ are evaluated at which grid points in our scheme.

To actually solve finite difference Equation \eqref{fdssaeqn1} requires an ``outer'' nonlinear iteration and an ``inner'' linear iteration.  The outer iteration produces a sequence of sparse linear problems in which the values of $\bar \nu H$ are (temporarily) known.  The inner iteration solves the linear problem for $v_1$ and $v_2$.

The outer iteration continues until successive values of $\bar \nu H$ are within a specified tolerance.  The criterion used for the results here is that the $L^2$ norm of the difference of successive values of $\bar \nu H$ is less than $10^{-4}$ of the $L^2$ norm of $\bar \nu H$ itself.  We do not require convergence of the velocities themselves in the outer iteration.

The inner linear problems can each be thought of as ``$A\bv=\bb$'', with the approximated driving stress $\rho g H \grad h$ as $\bb$.  The membrane stresses and the basal stress $\tau_b$ all appear on the left side as parts of the matrix $A$.  We factor $\tau_{b}$ into a coefficient, which depends in a strongly nonlinear way on the velocity, times the velocity.

If the grid has $M_1\times M_2$ horizontal points then there are $2 M_1 M_2$ scalar unknowns.  Using $M_1=M_2=301$ as in the 5 km grid results shown, each linear system has $1.8\times 10^5$ rows.  Five grid values of $v_1$ and eight grid values of $v_2$ are involved in Equation \eqref{fdssaeqn1}, so there are 13 nonzero coefficients per row.  These sparse linear problems are solved using a parallel iterative solver within PETSc \citep{petsc-user-ref}.  For the results here we used the PETSc default, GMRES(30) with block Jacobi and ILU preconditioning \citep{Saad}.  This solver/preconditioner pair is just one choice of many ``Krylov'' iterative methods within PETSc, and other combinations work.  We use the default Krylov relative tolerance of $10^{-5}$.

\section{Traditional SIA sliding laws and their difficulties}  Consider a slab on a slope \citep{Paterson} of thickness $H_0$ and slope $\alpha$, with horizontal coordinate $x$, positive-upward vertical coordinate $z$, and horizontal velocity $u$.

Sliding laws of the type used in EISMINT II experiment H \citep{EISMINT00} and ISMIP-HEINO \citep{GreveTakahamaCalov} give the sliding velocity as a function of the driving stress when the basal ice reaches the pressure melting temperature $T_0^*$.  Suppose that the basal temperature $T$ increases along the flow, reaching pressure melting temperature $T_0^*$ at some location $x_0$, as shown in Figure \ref{fig:slabjump}.  Suppose $T< T_0^*$ for $x<x_0$ while $T=T_0^*$ for $x>x_0$.
  Concretely, suppose $u_b=0$ when $T< T_p$ and $u_b = C \rho g H_0 \alpha$ when $T = T_0^*$, for a positive constant $C$.  The horizontal velocity at any location $(x,z)$ in the ice is
	$$u(x,z) = u_b(x) + \int_{b(x)}^{z} \ddz{u}\,d\zeta.$$
The SIA computes $\partial u/\partial z$ from the ice temperature, surface slope, and depth below the surface.  It follows that there is a jump in the basal velocity which implies a jump in horizontal velocity at every vertical level $z$ between the base and the surface of the ice:
	$$u(x_0^+,z)-u(x_0^-,z) = C \rho g H_0 \alpha.$$

On the other hand, the ice is incompressible.  \emph{Therefore the vertical velocity is formally infinite at all points in this ice column.}  Indeed, the vertical velocity is 
	$$w(x,z) = - \int_{b(x)}^z \ddx{u} \,d\zeta,$$ 
at any place where the horizontal velocity is differentiable in $x$.  In the ice column at $x=x_0$ we necessarily interpret ``$\partial u/\partial x$'' as infinity.  (We have made harmless small-slope approximations, including assuming $w=0$ at the base of the ice.)

It remains to explain why this problem has not already stopped this ice sheet modeling practice.  Consider a numerical scheme in which $x_0$ is between a pair of gridpoints $x_-$ and $x_+$ which are separated by $\Delta x$.  Suppose $H_0=2000$ m, $\alpha = 0.001$, $\rho=910$ $\text{kg}\,\text{m}^{-3}$, $g=9.81$ $\text{m}\,\text{s}^{-2}$, and $C=10^{-3}\,\text{m}\,\text{Pa}^{-1}\text{a}^{-1}$ \citep[c.f.][]{EISMINT00}.  If 
	$$\ddx{u}(x_0,z) \approx \frac{u(x_+,z) - u(x_-,z)}{\Delta x}$$
then Table \ref{tab:explodingvert} gives estimates of the surface value of vertical velocity, starting with common values of $\Delta x$ used in ice sheet modeling, and then down to 1 km.  We see that for the rough grids used in EISMINT and ISMIP-HEINO, for example, this ``infinity'' is well-hidden unless the modeler carefully examines locations of slightly anomolous vertical velocities.  The problem starts appearing at the level of grid refinement which is necessary to resolve the stresses in individual ice streams, however.

Because the vertical velocity is a part of the energy conservation Equation \eqref{conserve}, however, anomolous vertical velocity is merely the start of unreasonable, time-dependent, thermomechanically-coupled behavior.  We hypothesize that the strange spokes in EISMINT II experiment H \citep{EISMINT00} are a consequence of the mechanism described in this appendix.  By contrast, the spokes for the nonsliding experiments in EISMINT II came from a different thermomechanical fluid instability mechanism, as discussed in \citep{BBL} and references.

We also hypothesize that the ``Heinrich events'' demonstrated in ISMIP-HEINO \citep{GreveTakahamaCalov} are consequences of thermomechanical coupling to this anomolous mechanism here.  This comment does not discount the possibility that ice sheet sliding is connected to Heinrich events.  Rather, we believe that continuum models for Heinrich events should not use SIA sliding laws of the type described in this Appendix.  Indeed, membrane stresses modulate sliding so the need to be included in any model for time-dependent ice stream behavior.

\begin{acknowledgments}
This paper was guided by correspondence and conversations with D.~Maxwell, S.~Price, M.~Truffer, and C.~Schoof.  The major influence of the work of C.~Schoof should be clear.  Encouragement and advice from C.~Lingle was critical to developing PISM.  The Arctic Regional Supercomputer Center (ARSC) has provided more than 50k processor hours of supercomputer time, allowing us to recover from our mistakes, and Figure \ref{fig:timingP1parallel} came from runs on ARSC's Cray XT5.\end{acknowledgments}


\end{article}

\clearpage\newpage


\begin{table}
\caption{Notation, units, and values for constants.  Vectors are in \textbf{bold}.}\label{tab:notation}
\begin{tabular}{llll}
Symbol & Description & SI Units & Value  \\ \hline
 & seconds per year & & 31556926 \\
 $\grad$, $\Div$ & gradient and divergence in horizontal coordinates & $\text{m}^{-1}$ & \\
$A(T^*)$ & temperature-dependent rate factor in flow law; Eqn \eqref{constitutive} & $\text{Pa}^{-3}\,\text{s}^{-1}$  & \\
$B(T^*)$ & ice hardness; $B(T^*) = A(T^*)^{-1/n}$ & Pa $\text{s}^{1/3}$ & \\
$b$ & bed elevation & m & \\
$\beta$ & dependence of melting point on depth & $\text{K}\,\text{m}^{-1}$ & $8.66 \times 10^{-4}$ \\
$c_i$ & specific heat capacity for ice  & $\text{J}\, (\text{kg}\;\text{K})^{-1}$ & $2009$\\
$D_{ij}$ & strain rate tensor & $\text{s}^{-1}$ & \\
$D_L$ & depth of lithosphere thermal model & m & 515 \\
$D_{\text{SIA}}$ & diffusivity associated to SIA; Eqn \eqref{DSIA} & $\text{m}^2\,\text{s}^{-1}$ \\
$\delta$ & regularizing velocity for plastic till; Eqn \eqref{plasticbasereg} & $\text{m}\,\text{s}^{-1}$ & $0.01\,\text{m}\,\text{a}^{-1}$ \\
$\eps$ & regularizing velocity for effective viscosity; Eqn \eqref{barnureg} & $\text{m}\,\text{s}^{-1}$ & $1.0\,\text{m}\,\text{a}^{-1}$ \\
$f(|\bv|)$ & weighting function for combining velocities; Eqns \eqref{superAverage}, \eqref{fofv} & & \\
$\phi$ & till friction angle; Eqn \eqref{MohrCoulomb} &  & \\
$G_0$ & geothermal flux  & $\text{W}\,\text{m}^{-2}$ & $.042$\\
$g$ & acceleration of gravity  & $\text{m}\,\text{s}^{-2}$ & $9.81$\\
$h$ & ice surface elevation & m & \\
$H$ & ice thickness & m & \\
$k_i$ & thermal conductivity of ice & $\text{W}\,(\text{K}\;\text{m})^{-1}$ & 2.10 \\
$K_r$ & thermal diffusivity of lithosphere & $\text{m}^2\,\text{s}^{-1}$ & $9.09 \times 10^{-7}$ \\
$L$ & latent heat of fusion for ice & $\text{J}\,\text{kg}^{-1}$ & $3.35\times 10^5$ \\
$\tilde L$ & diffusion distance for melt water thickness; Eqn \eqref{Hmelt} & m & 20 km \\
$L_\nu$ & regularizing length for effective viscosity; Eqn \eqref{barnureg} & m & $10^6$ \\
$M$ & ice-equivalent surface mass balance ($M>0$ is accum.) & $\text{m}\,\text{s}^{-1}$ & \\
$n$ & Glen exponent  & & 3\\
$\bar\nu$ & vertically-averaged (effective) viscosity in SSA; Eqn \eqref{barnu} & Pa s & \\
$\bQ$ & horizontal ice flux & $\text{m}^2\,\text{s}^{-1}$  & \\
$p$ & pressure (in ice) & Pa & \\
$p_w$ & pore water pressure in till; Eqn \eqref{porewatermodel} & Pa & \\
$R$ & gas constant  & $\text{J}\,(\text{mol}\;\text{K})^{-1}$ & 8.314 \\
$\rho_i$ & density of ice  & $\text{kg}\,\text{m}^{-3}$ & 910 \\
$S$ & ice-equivalent basal mass balance ($S>0$ is melting) & $\text{m}\,\text{s}^{-1}$ & \\
SIA & ``shallow ice approximation'', esp.~Eqn \eqref{velocitySIA} & & \\
SSA & ``shallow shelf approximation'', esp.~Eqn \eqref{SSA} & & \\
$T$ & absolute ice (or bedrock) temperature & K & \\
$T^*$ & pressure corrected ice temperature; $= T_0 - \beta(h-z)$ & K & \\
$T_0$ & melting temperature for ice & K & $273.15$ \\
$T_s$ & surface temperature & K & \\
$t$ & time & s & \\
$\tilde t$ & diffusion time for melt water thickness; Eqn \eqref{Hmelt} & s & 1000 a \\
$\tau_{b,i}$ & basal shear stress applied to ice; Eqn \eqref{plasticbase} & Pa & \\
$\tau_c$ & till yield stress & Pa & \\
$\tau_{ij}$ & deviatoric stress tensor & Pa &  \\
$\bu=(u_1,u_2,u_3)$ & velocity computed from non-sliding SIA; Eqn \eqref{velocitySIA} & $\text{m}\,\text{s}^{-1}$ & \\
$\bU=(U_1,U_2,U_3)$ & ice velocity; note Eqn \eqref{superAverage} & $\text{m}\,\text{s}^{-1}$ & \\
$\bar \bU$ & vertically-averaged horizontal velocity & $\text{m}\,\text{s}^{-1}$ & \\
$\bU_b$ & basal value of the ice velocity & $\text{m}\,\text{s}^{-1}$ & \\
$\bv=(v_1,v_2,v_3)$ & velocity computed from plastic till SSA; Eqn \eqref{SSA} & $\text{m}\,\text{s}^{-1}$ & \\
$\Hmelt$ & effective thickness of stored basal water; Eqn \eqref{Hmelt} & m & \\
$\Hmelt_0$ & limit on $\Hmelt$; basal drainage parameter & m & 2.0 \\
$x_1,x_2,x_3$ & cartesian coordinates & m &  \\
$z$ & alternate notation for $x_3$; positive upward & m &  \\
\end{tabular}
\end{table}

\begin{table}
\caption{Constants for the exact SSA solution which is equation 4.3 in \citep{SchoofStream}, with values $B,f$ from the same source.}\label{tab:testIconst}
\begin{tabular}{lll}
Symbol & Meaning & Value\\ \hline
$B$ & ice hardness & $3.7\times 10^8\, \text{Pa} \,\text{s}^{1/3}$  \\
$f$ & scale for till yield stress & $17.854$ kPa \\
$h_0$ & constant ice thickness & 2000 m \\
$L$ & half-width of weak till & 40 km \\
$m$ & power & 10 \\
$\tan\theta$ & bed slope & $0.001$ \\
\end{tabular}
\end{table}

\begin{table}
\caption{Parameter studies:  Each experiment runs for 5000 models years on grids of $15$, $10$, $7.5$, and $5$ km spacing, starting from the same initial state.}\label{tab:experiments}
\begin{tabular}{lll}
Name & Parameter explored & Comments \\ \hline
P1 & width of weak till strip & \emph{10 km grid run extended to 100k years} \\
P2 & orientation of weak till strip & \emph{only three weak till strips} \\
P3 & bed slope & \\
P4 & strength of downstream till & \\
\end{tabular}
\end{table}

\begin{table}
\caption{Numerical surface value of the vertical velocity from a traditional SIA sliding law.}\label{tab:explodingvert}
\begin{tabular}{c|ccccccc}
$\Delta x$ (km) & 40 & 25 & 15 & 10 & 5 & 2 & 1 \\ \hline
surface value of $w$ (m/a) & -0.9 & -1.4 & -2.4 & -3.6 & -7.1 & -17.9 & -35.7 
\end{tabular}
\end{table}



\newcommand{\tfinclude}[1]{\includegraphics[width=11pc]{tillphiP#1}}
\begin{figure}
\mbox{a\,\tfinclude{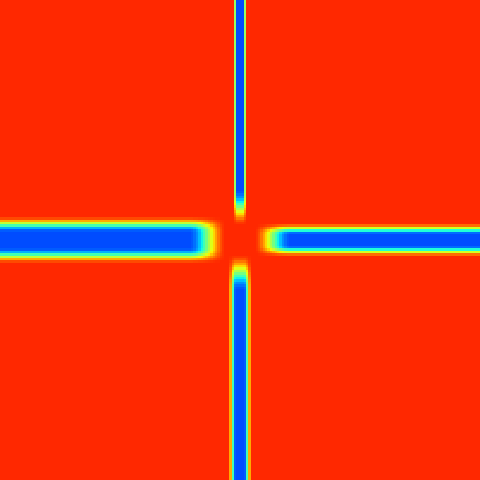}\quad b\,\tfinclude{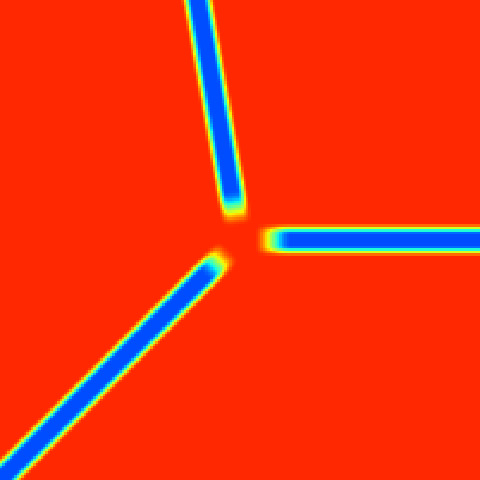}\quad c\,\tfinclude{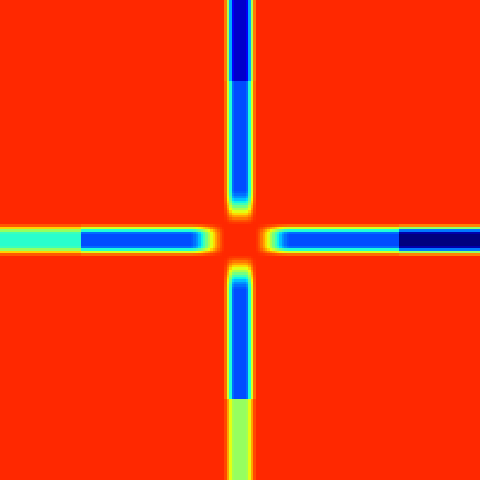}}
\caption{Maps of till friction angle $\phi(x,y)$ for experiments P1 (a), P2 (b), and P4 (c).  Red is $15^\circ$.  Blue in weak strips in a,b is $5^\circ$, also in upstream parts of weak strips in c.  See text for downstream $\phi$ values in experiment P4 and for description of experiment P3.}
\label{fig:tillmaps}
\end{figure}

\begin{figure}
\noindent\includegraphics[width=20pc]{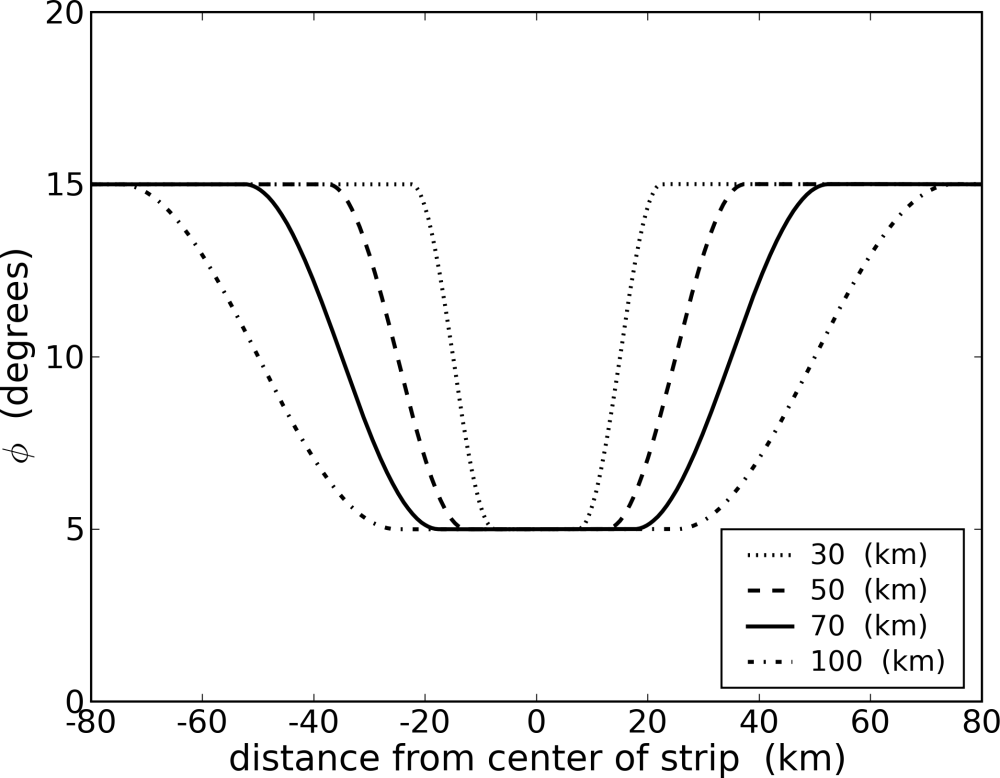}
\caption{Till friction angle $\phi$ across the four weak till strips in experiment P1, with given strip widths.}
\label{fig:crossprofileP1phi}
\end{figure}

\begin{figure}
\noindent\includegraphics[width=20pc]{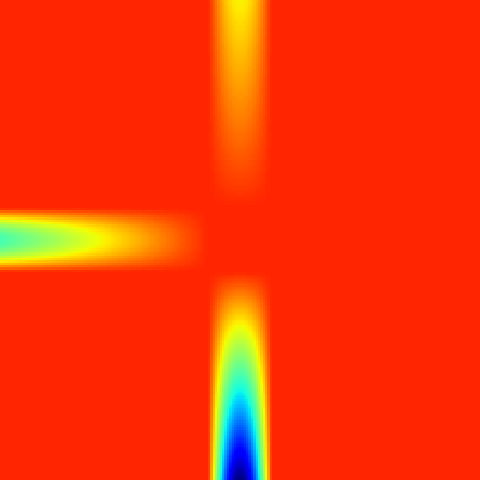}
\caption{Trough bed topography in experiment P3.  Red is flat plateau, and troughs descend $0$ m (east), $500$ m (north), $1000$ m (west), and $2000$ m (south) below the plateau.}
\label{fig:P3bed}
\end{figure}

  
\begin{figure}
\noindent\includegraphics[width=20pc]{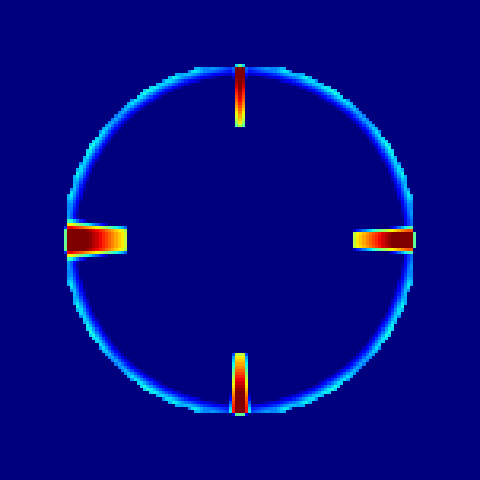}
\caption{Difference $\Delta\tau = \rho g H |\grad h|  - \tau_c$ between basal driving stress and till yield stress at start of experiment P1.  $\tau_c$ computed from Equation \eqref{MohrCoulomb}.  Darkest red regions have $\Delta\tau \ge 5 \times 10^4$ Pa, so they slide strongly at the beginning of the run.}
\label{fig:stresscompareP1start}
\end{figure}

\begin{figure}
\mbox{
\raisebox{1.7in}{
\begin{tabular}{c}
\mbox{
\includegraphics[width=2.0in,keepaspectratio=true]{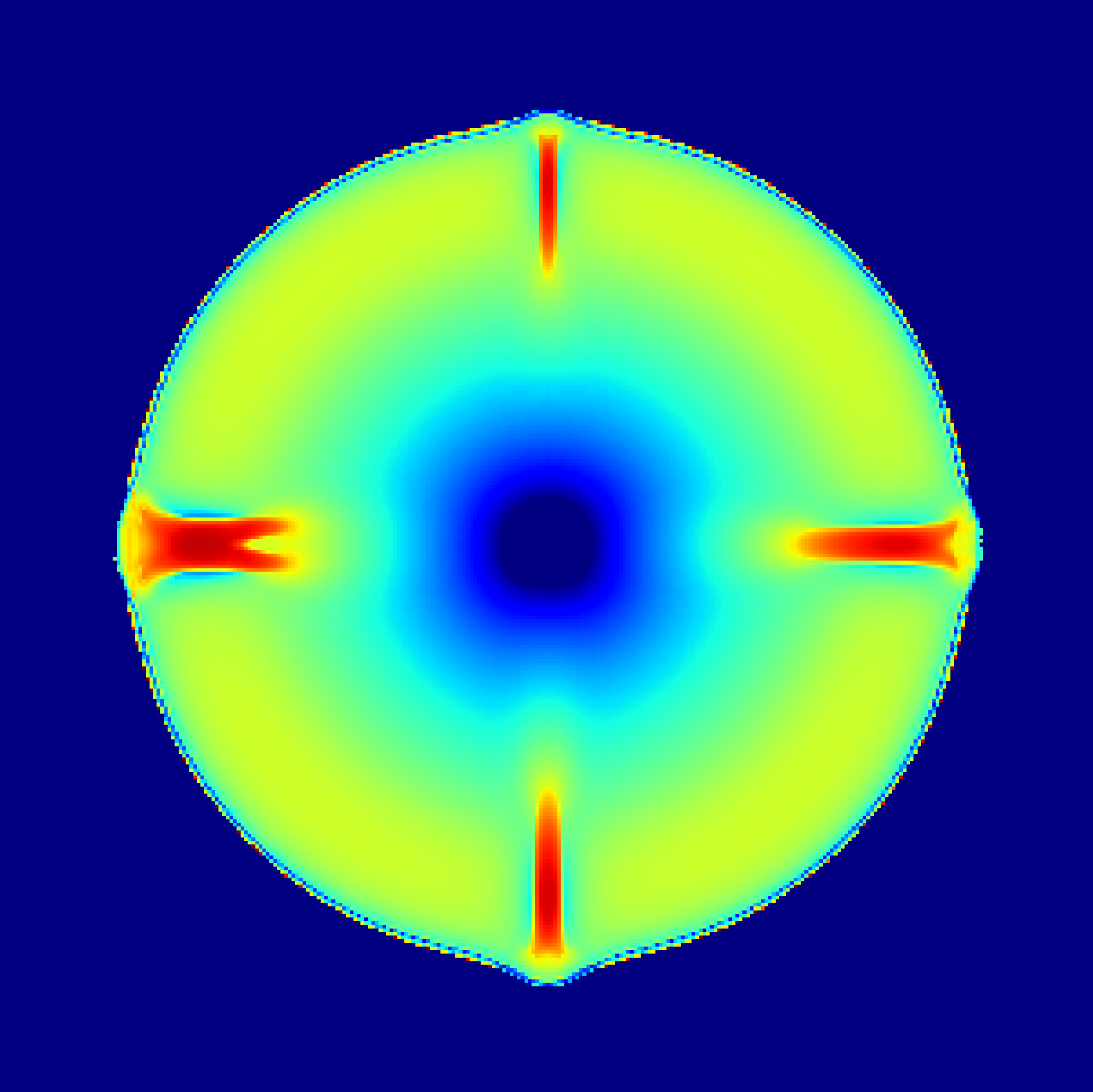}
\includegraphics[width=2.0in,keepaspectratio=true]{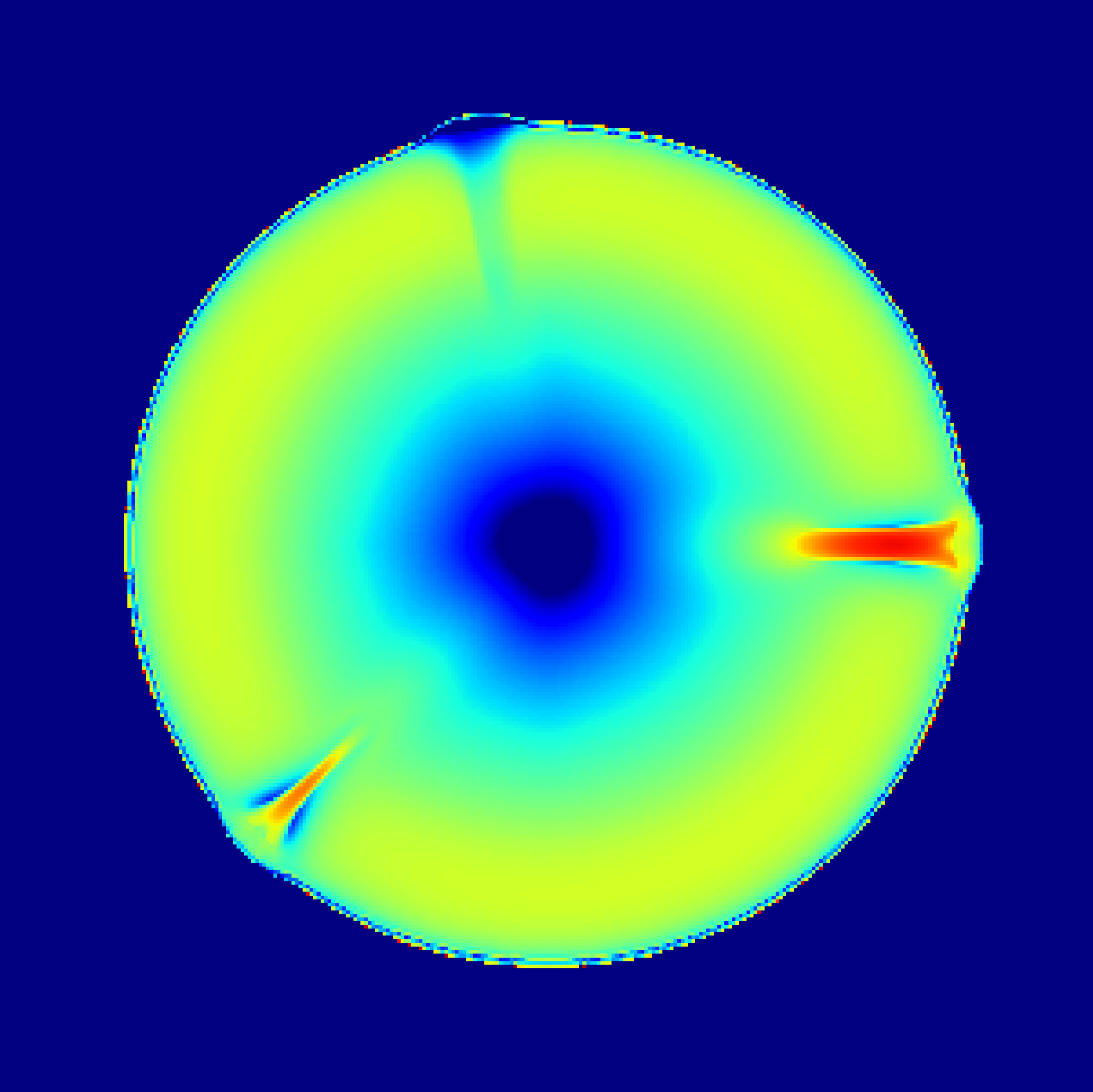}
}\\
\mbox{
\includegraphics[width=2.0in,keepaspectratio=true]{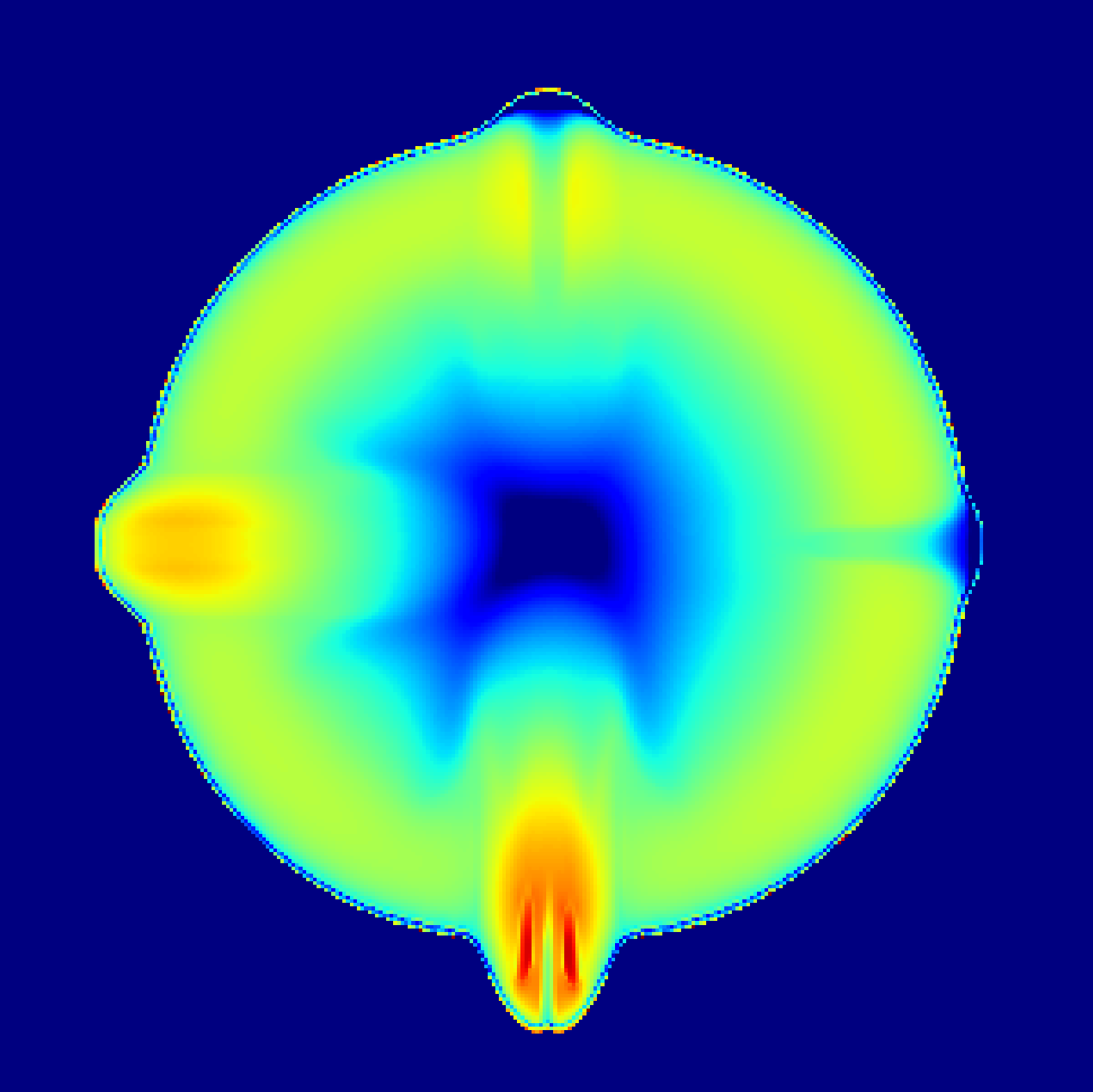}
\includegraphics[width=2.0in,keepaspectratio=true]{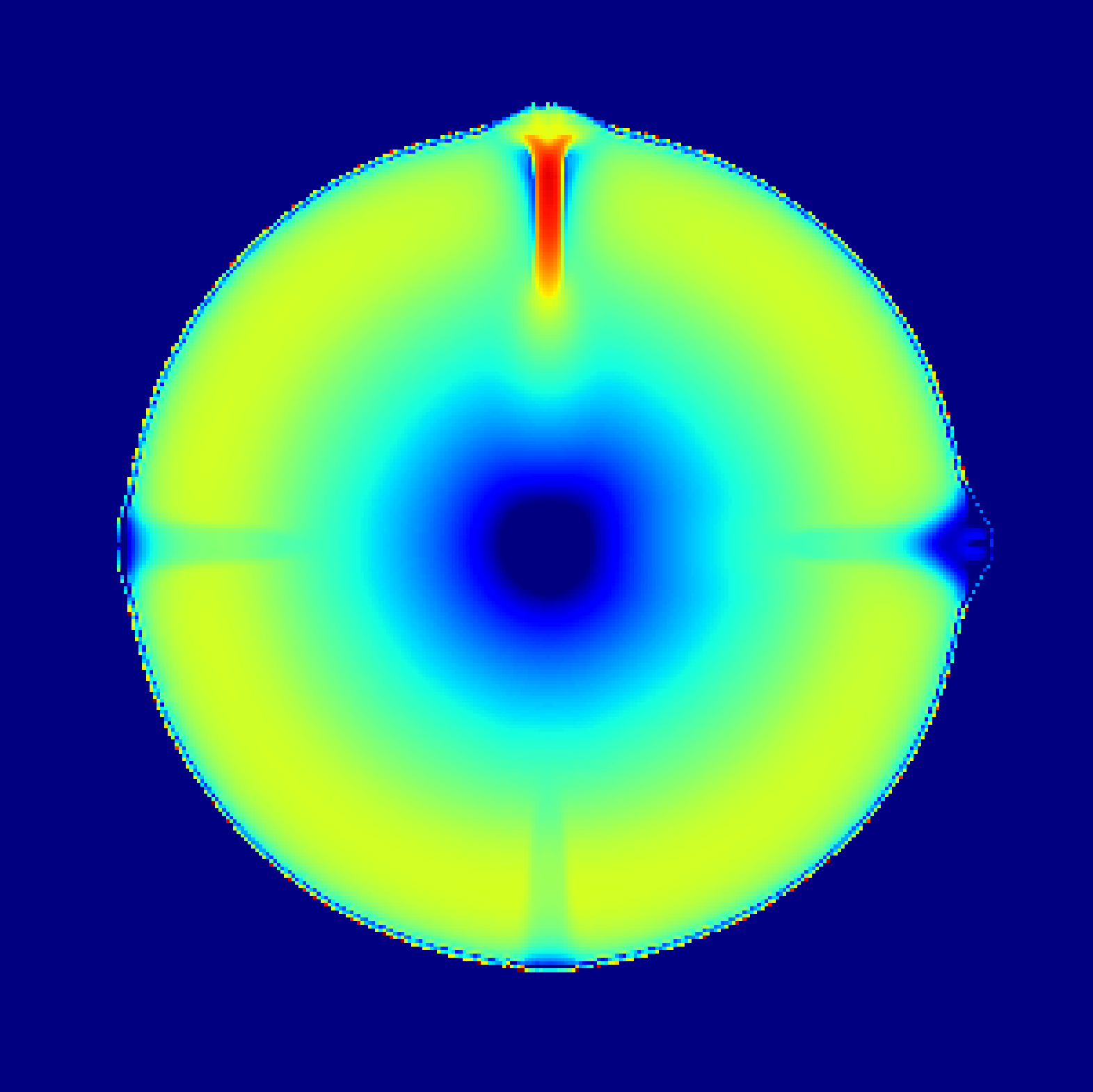}
}
\end{tabular}
}
\,
\includegraphics[height=3.5in,keepaspectratio=true]{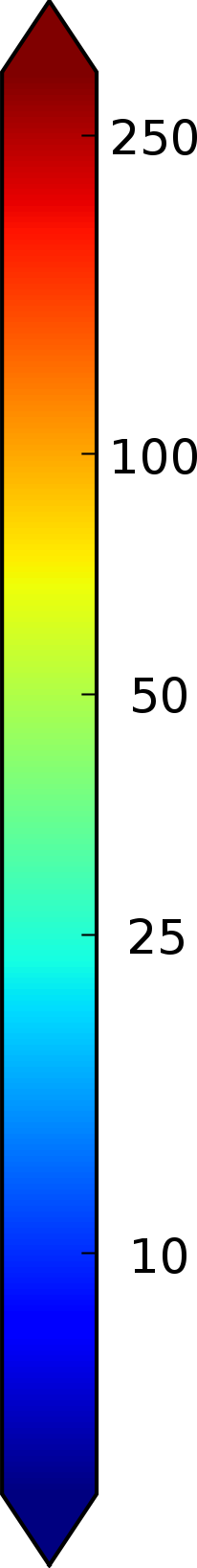}
}
\caption{Modeled horizontal surface ice speed (m/a) in logarithmic color scale, at 5 km resolution and at the end of 5 ka runs: experiments P1 (upper left), P2 (upper right), P3 (lower left), P4 (lower right).}
\label{fig:PallSpeed}
\end{figure}

\clearpage\newpage

\begin{figure}
\noindent\includegraphics[width=20pc]{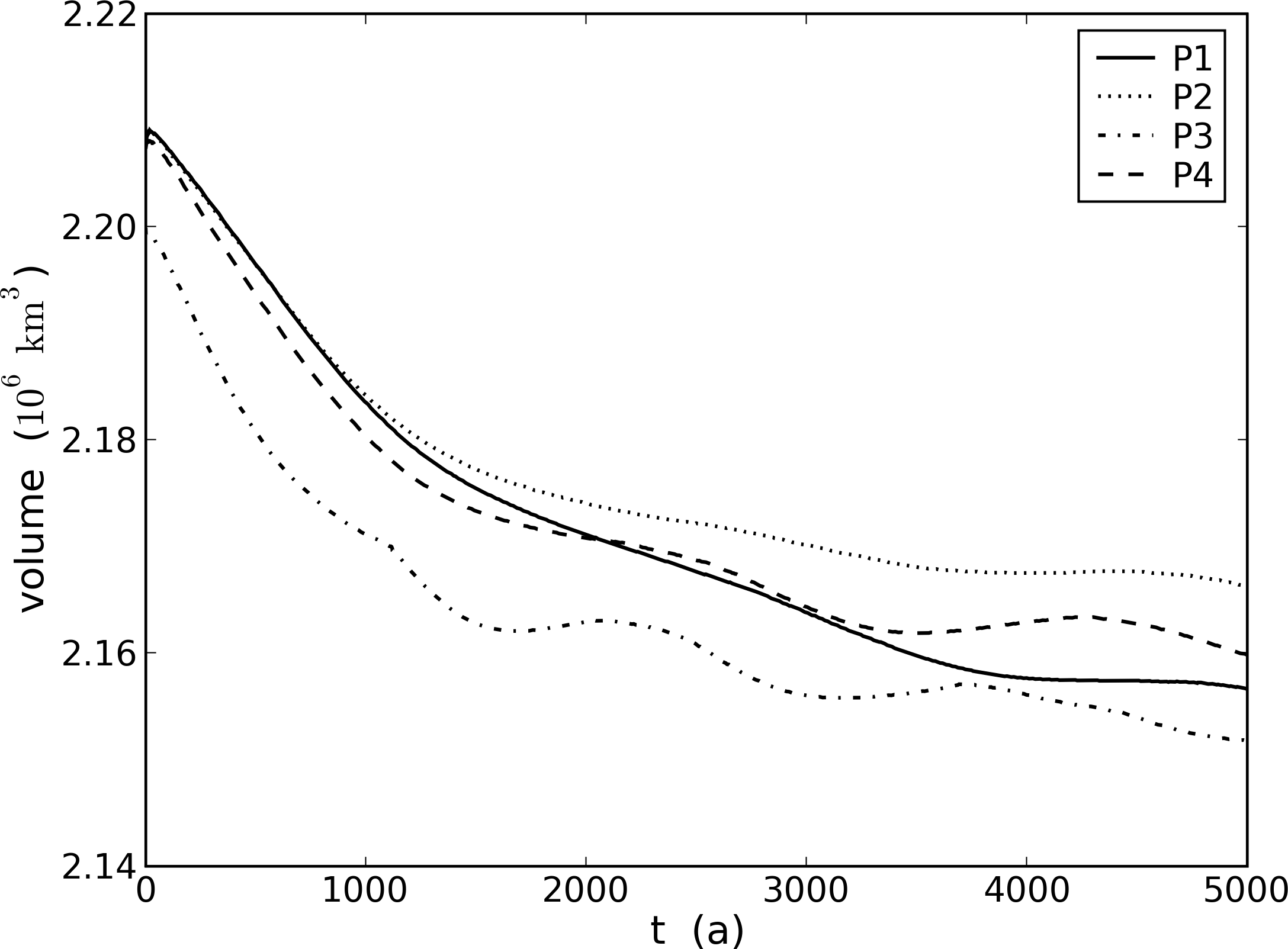}
\caption{Evolution of model ice sheet volume for the four experiments at 5 km grid resolution.}
\label{fig:PallVolTS}
\end{figure}

\begin{figure}
\noindent\includegraphics[width=20pc]{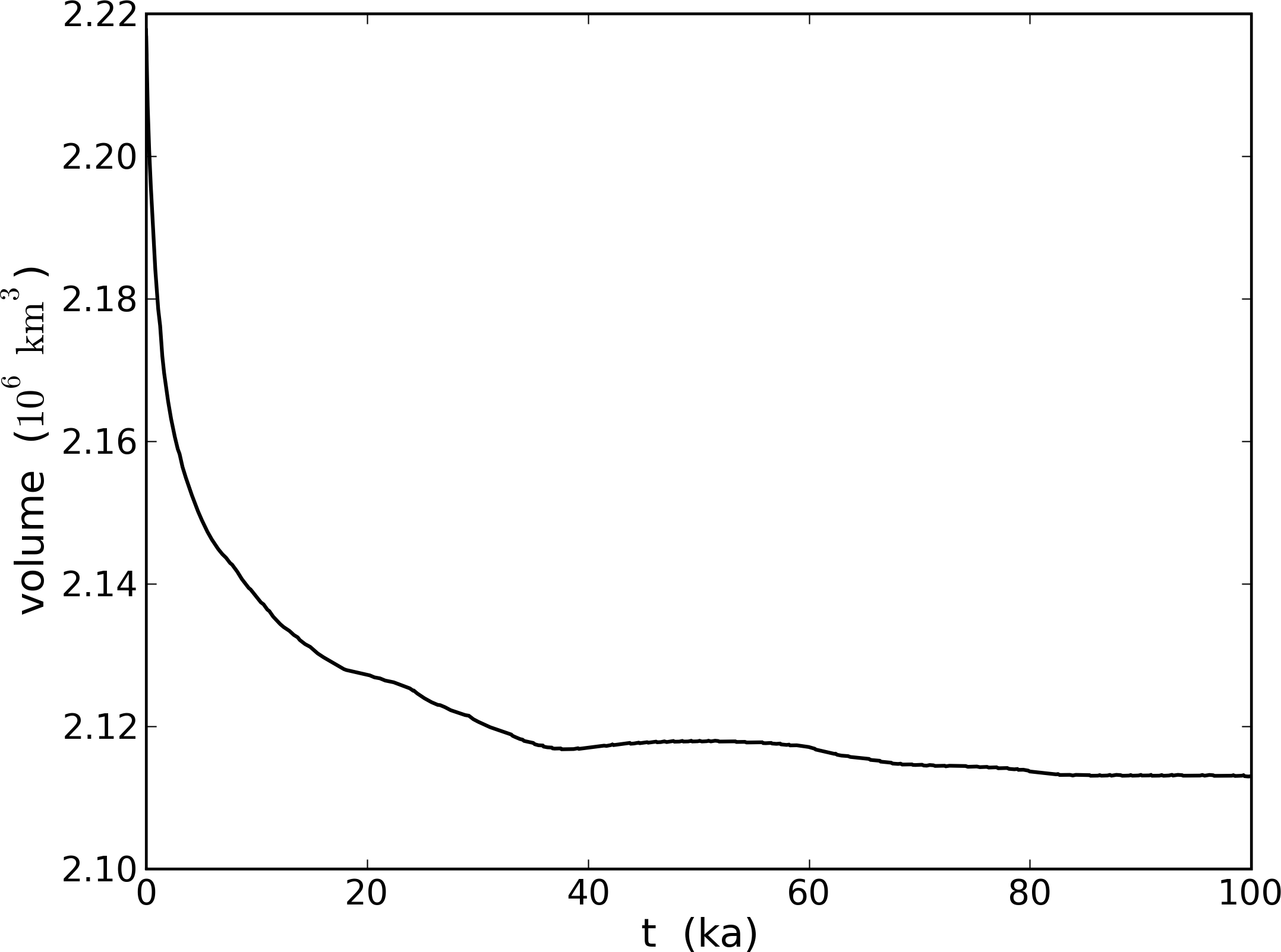}
\caption{Evolution of model ice sheet volume for experiment P1 at 10 km grid resolution over a 100 ka span.}
\label{fig:P1VolTScont}
\end{figure}

\begin{figure}
\noindent\includegraphics[width=20pc]{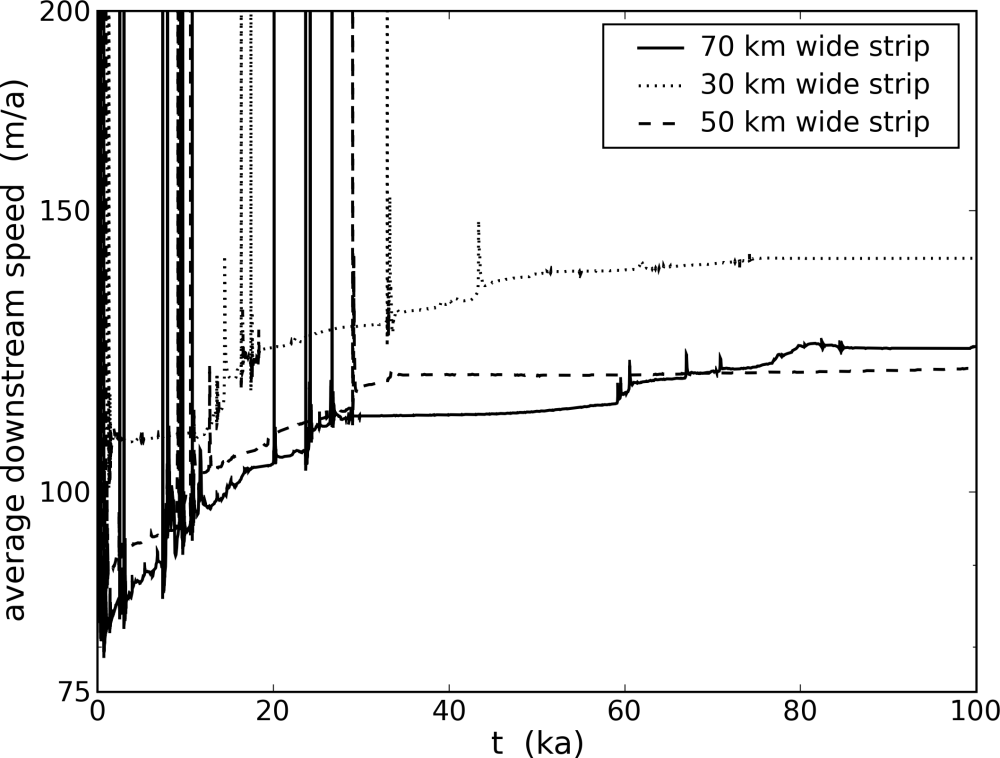}
\caption{Average downstream speed (magnitude of vertically-averaged horizontal velocity) for three of the four ice streams in experiment P1, over a 100 ka span.}
\label{fig:P1DownSpeedcont}
\end{figure}

\begin{figure}
\noindent\includegraphics[width=20pc]{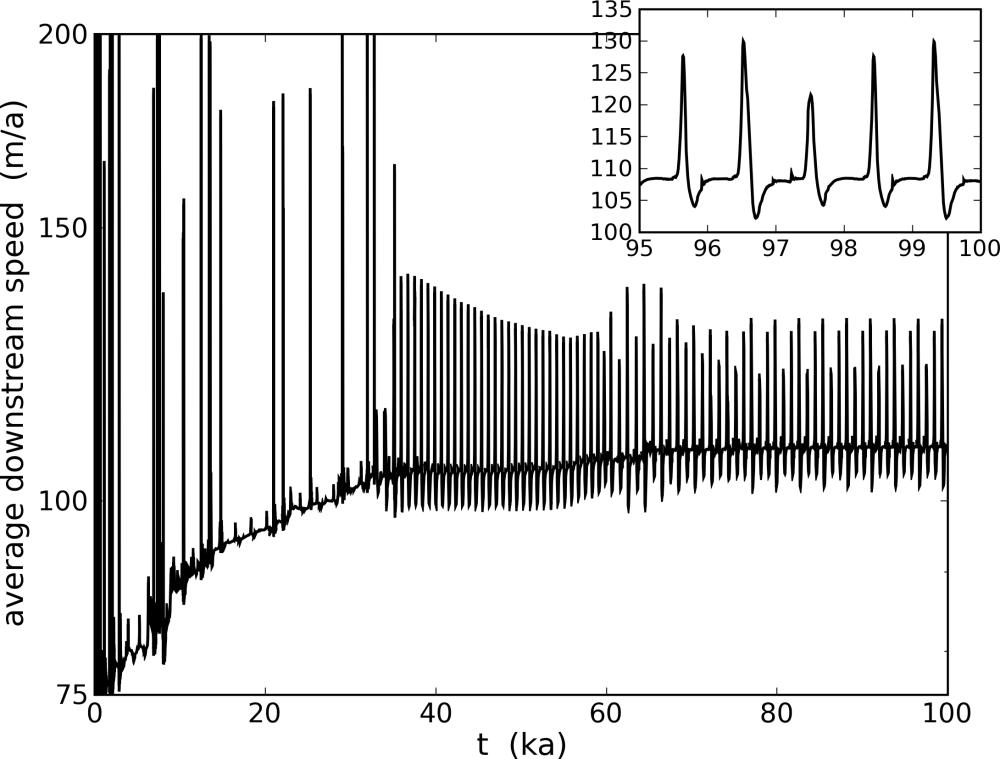}
\caption{Average downstream speed for the 100 km wide ice stream in experiment P1, over a 100 ka span.  Inset shows final 5 ka of run.}
\label{fig:P1DownSpeedcontWidth100}
\end{figure}

\begin{figure}
\noindent\includegraphics[width=20pc]{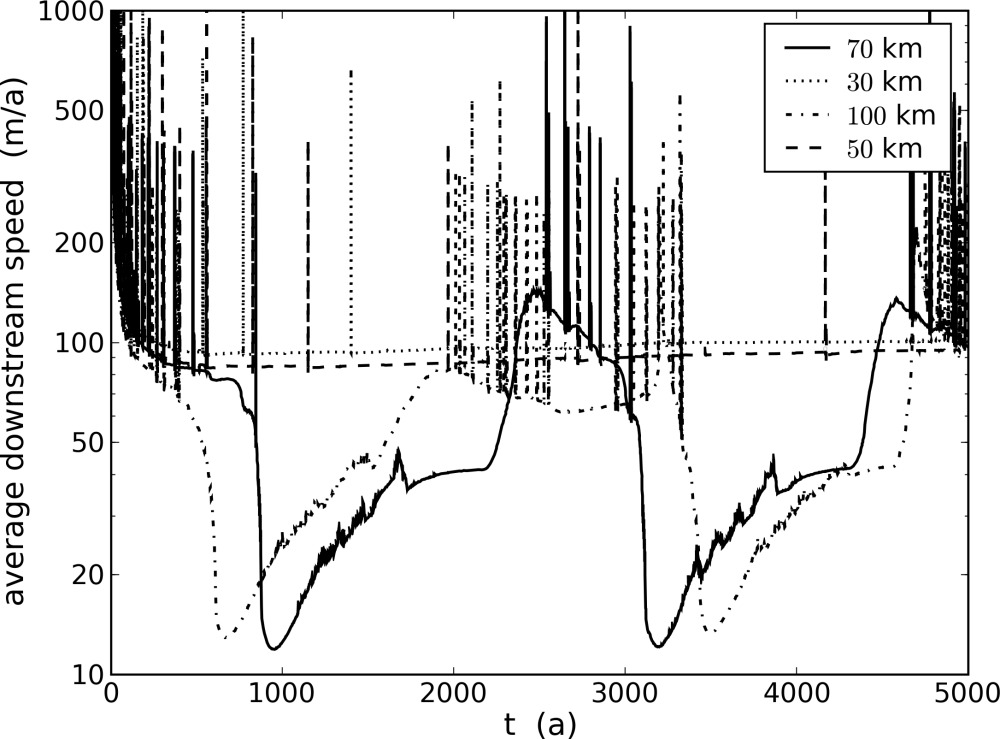}
\caption{Average downstream speed for the ice streams in experiment P1, at 5 km resolution for a 5 ka run.  Weak till strip width is given in the legend.}
\label{fig:P1DownSpeed}
\end{figure}

\begin{figure}
\noindent\includegraphics[width=20pc]{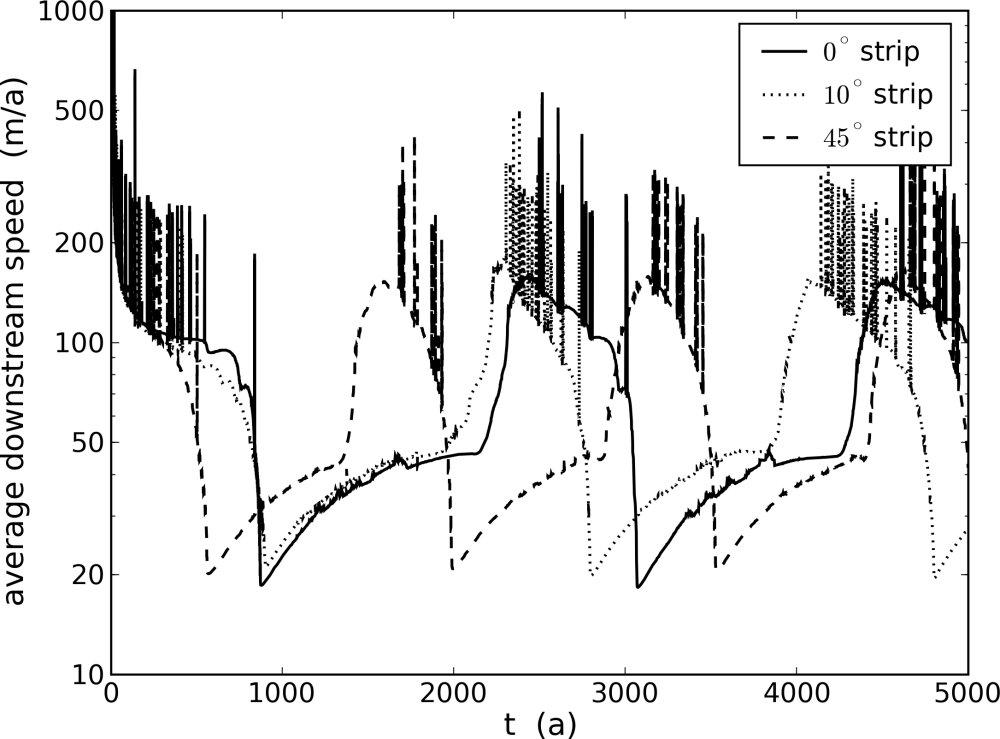}
\caption{Average downstream speed for the three ice streams in experiment P2, at 5 km resolution.  Strip angle, relative to the closest grid (cardinal) direction, is given in the legend.}
\label{fig:P2DownSpeed}
\end{figure}

\begin{figure}
\noindent\includegraphics[width=20pc]{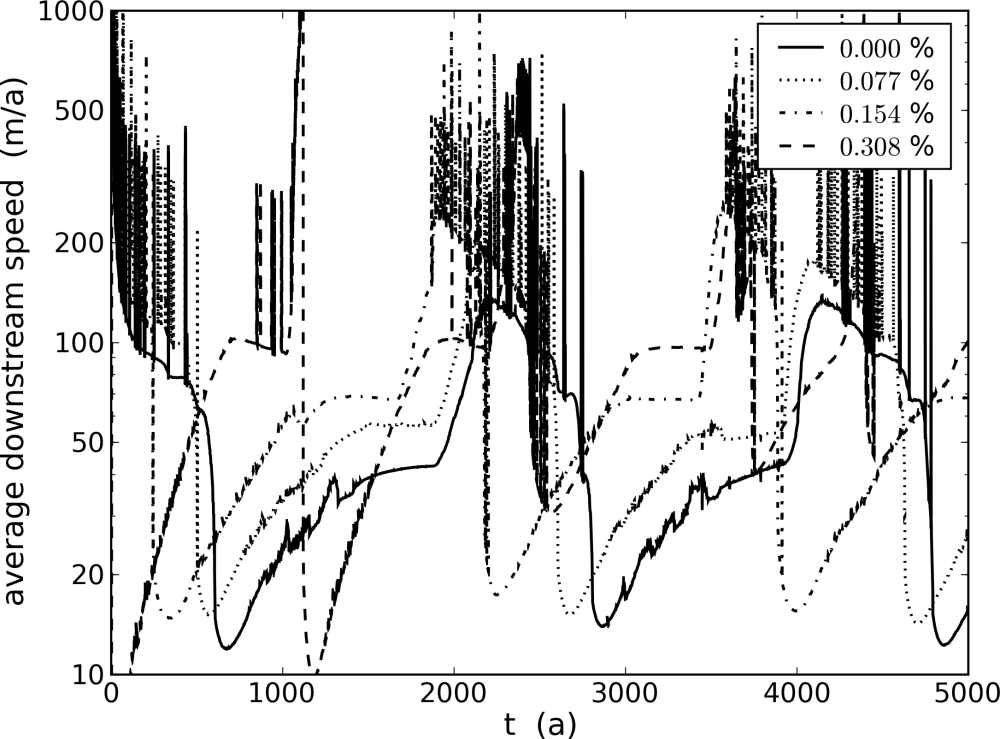}
\caption{Average downstream speed for the ice streams in experiment P3, at 5 km resolution.  Trough slope is given in the legend as a percentage.}
\label{fig:P3DownSpeed}
\end{figure}

\begin{figure}
\noindent\includegraphics[width=20pc]{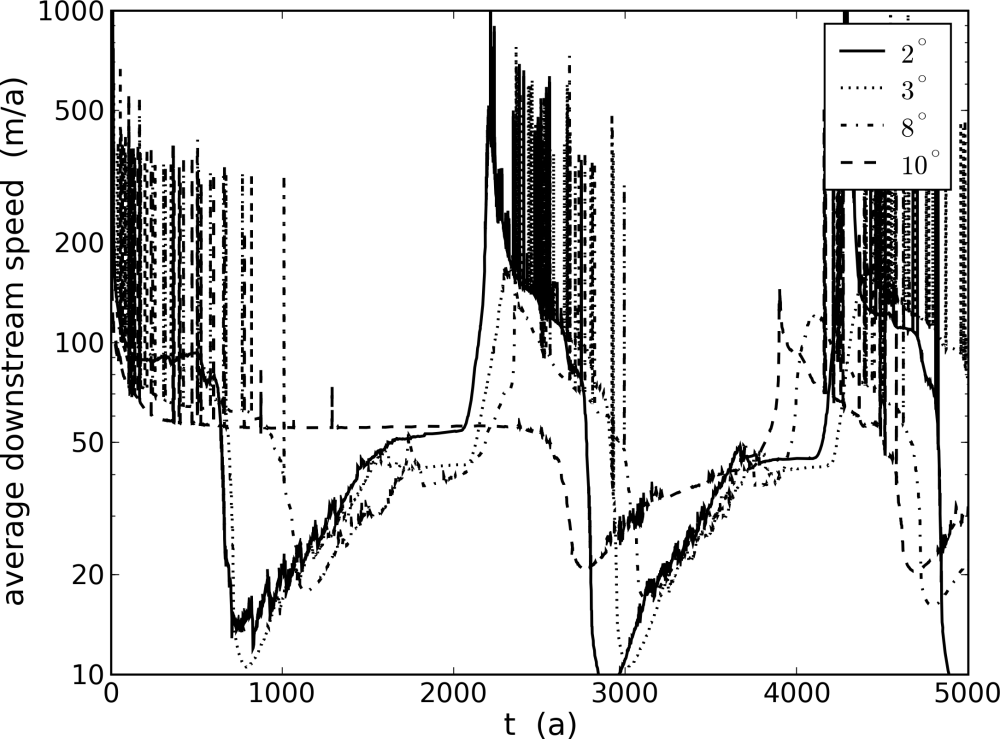}
\caption{Average downstream speed for the ice streams in experiment P4, at 5 km resolution.  Downstream till friction angle is given in the legend.}
\label{fig:P4DownSpeed}
\end{figure}


\begin{figure}
\noindent\includegraphics[width=20pc]{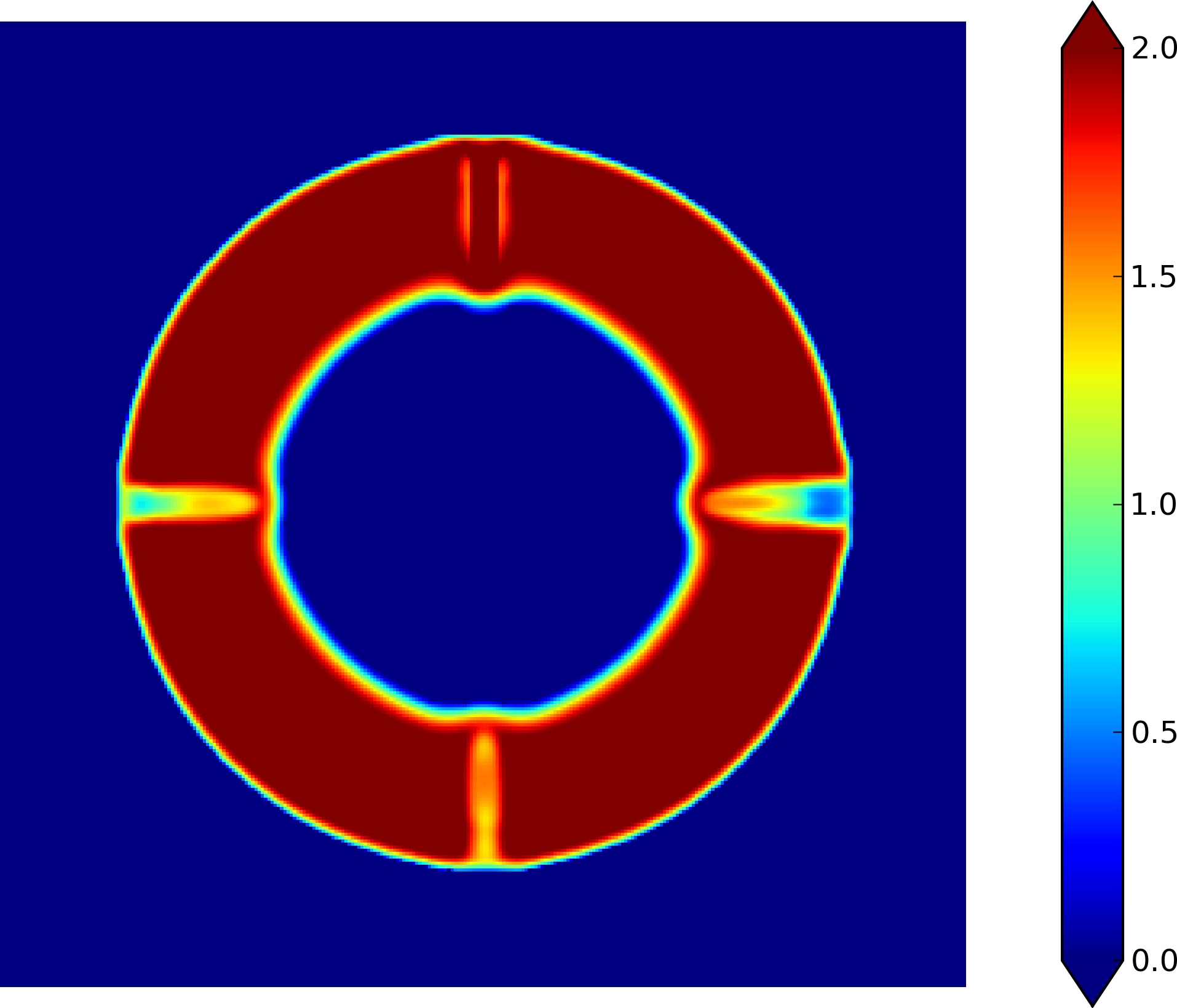}
\caption{Modeled basal water effective thickness (m) for experiment P4, at 5 km resolution and at the end of the 5 ka run.  The effective thickness is limited to a maximum of $2$ m in the model.}
\label{fig:P4bwat}
\end{figure}

\begin{figure}
\noindent\includegraphics[width=20pc]{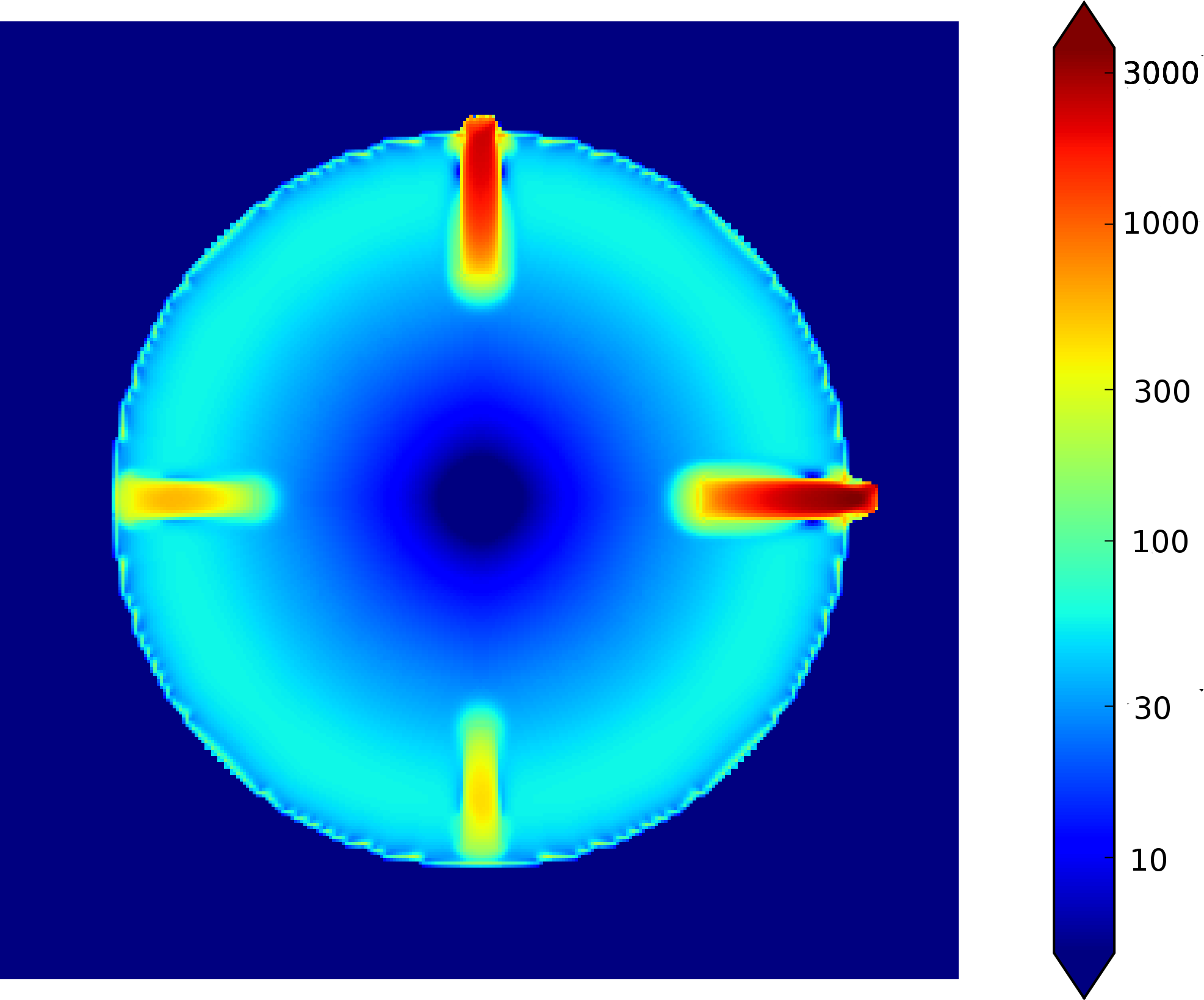}
\caption{Modeled horizontal surface ice speed (m/a) from experiment P4, at 5 km resolution but at only 5 model years from the start of the run.}
\label{fig:P4at5yr_csurf}
\end{figure}

\begin{figure}
\noindent\includegraphics[width=20pc]{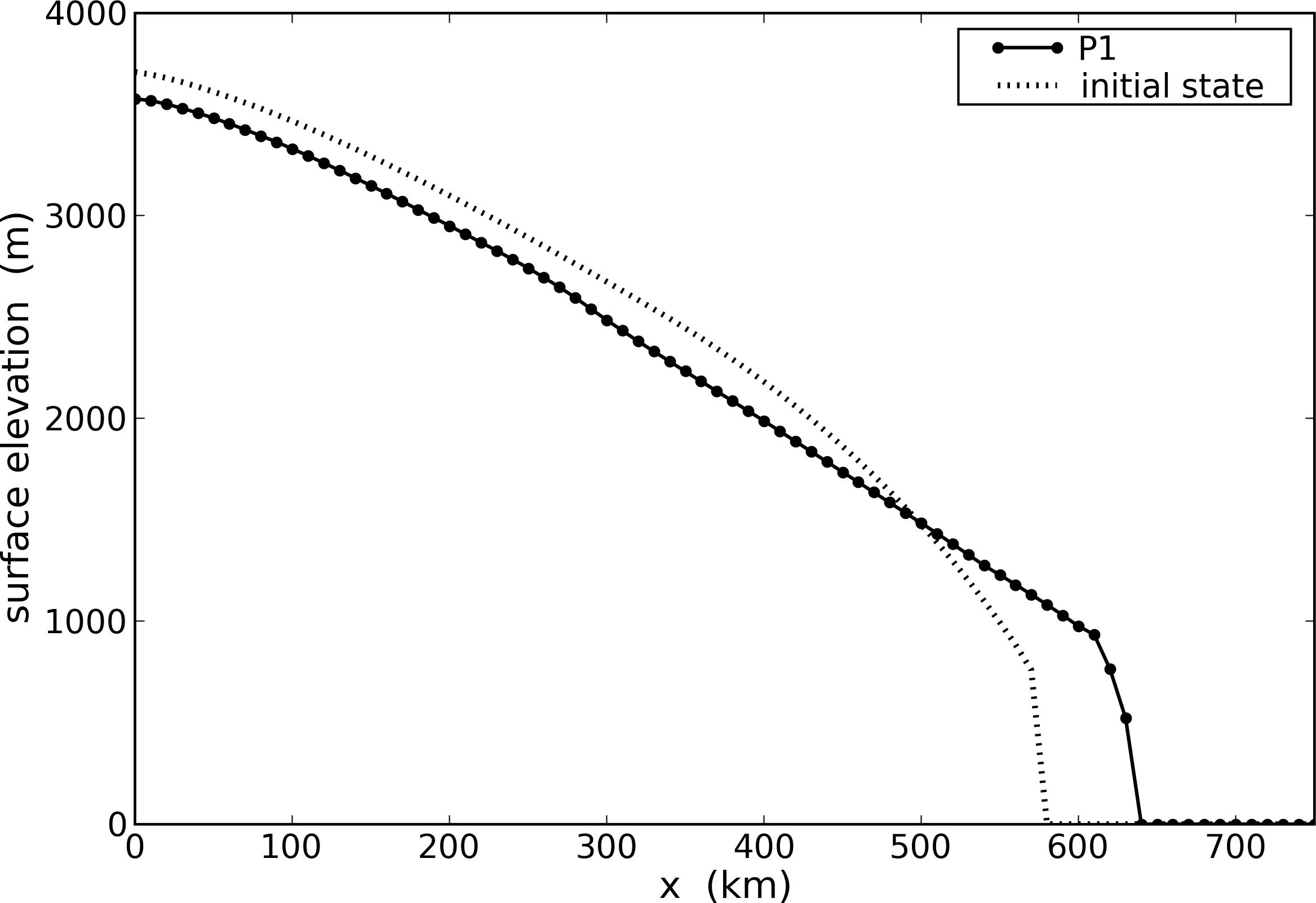}
\caption{Surface elevation, from the center of the ice sheet along the flow line corresponding to the center of the 70 km wide weak till strip in experiment P1, at 10 km resolution and at the end of the 100 ka run, compared to the initial state.}
\label{fig:P1vsP0A}
\end{figure}

\begin{figure}
\noindent\includegraphics[width=20pc]{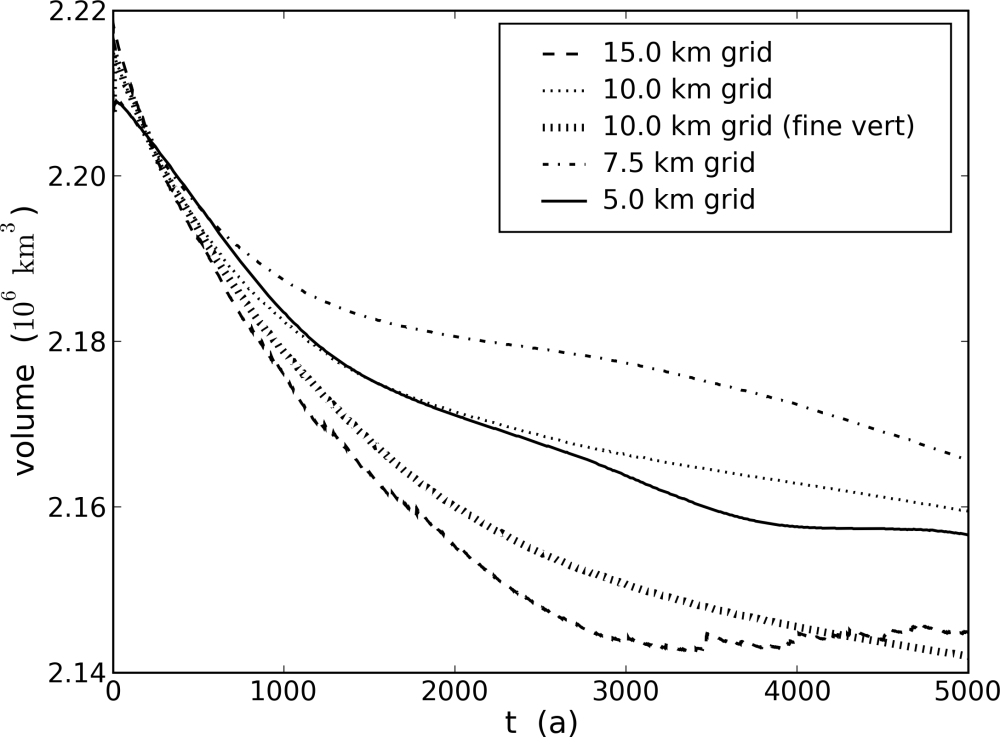}
\caption{Effect of grid refinement on the volume time series for the 5 ka run of experiment P1.}
\label{fig:gridP1vol}
\end{figure}

\begin{figure}
\noindent\includegraphics[width=20pc]{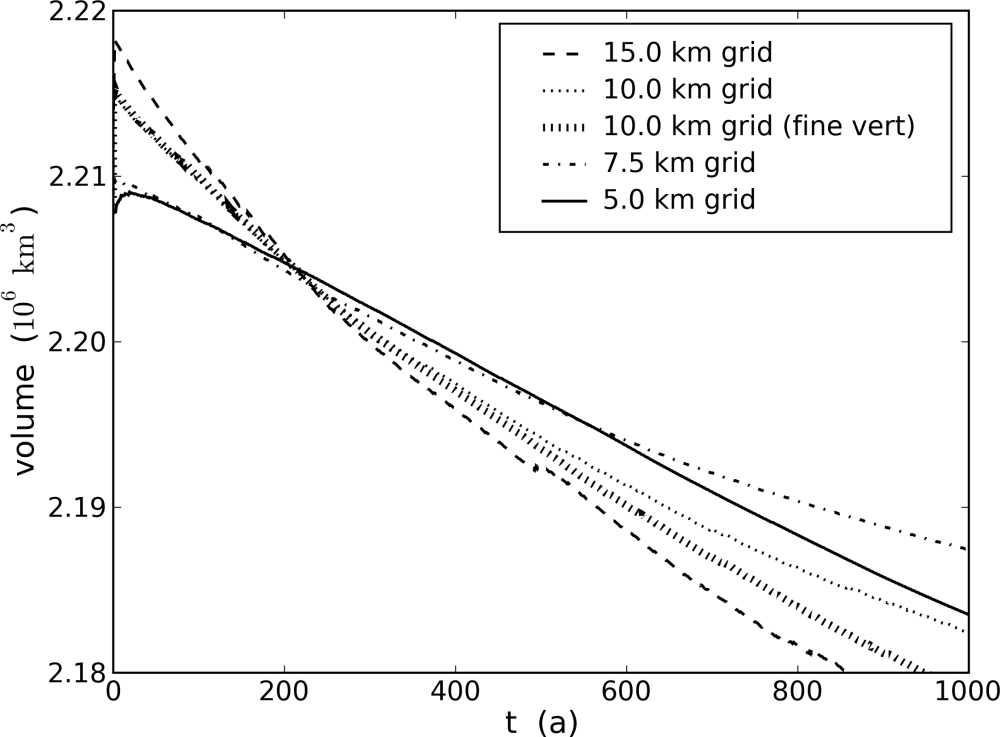}
\caption{Effect of grid refinement on the volume time series in first 1 ka (experiment P1).}
\label{fig:gridP1vol1ka}
\end{figure}

\begin{figure}
\noindent\includegraphics[width=20pc]{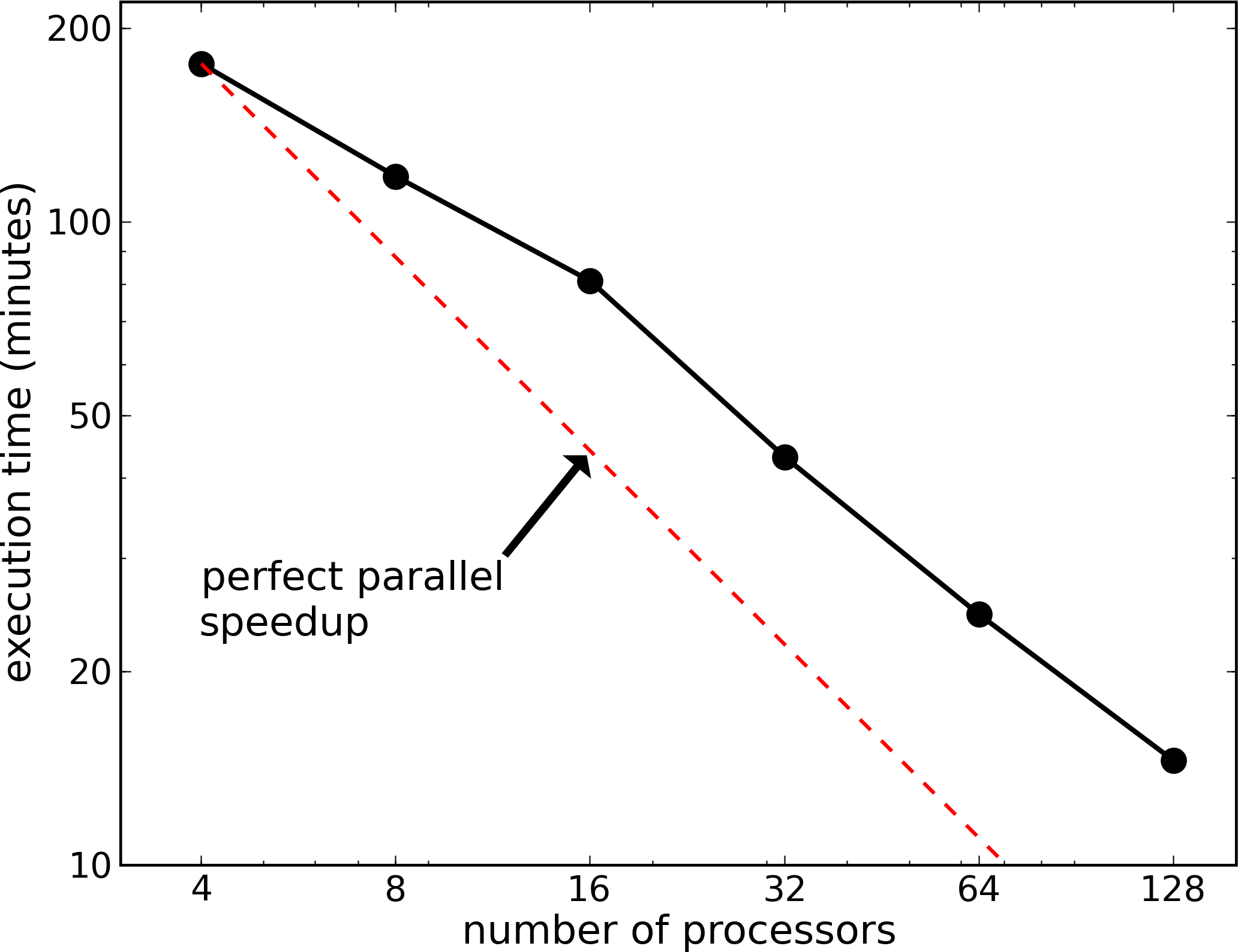}
\caption{Wall clock time for experiment P1 decreases steadily with more processors.}
\label{fig:timingP1parallel}
\end{figure}


\begin{figure}
\mbox{
\raisebox{1.7in}{
\begin{tabular}{c}
\includegraphics[width=2.5in,keepaspectratio=true]{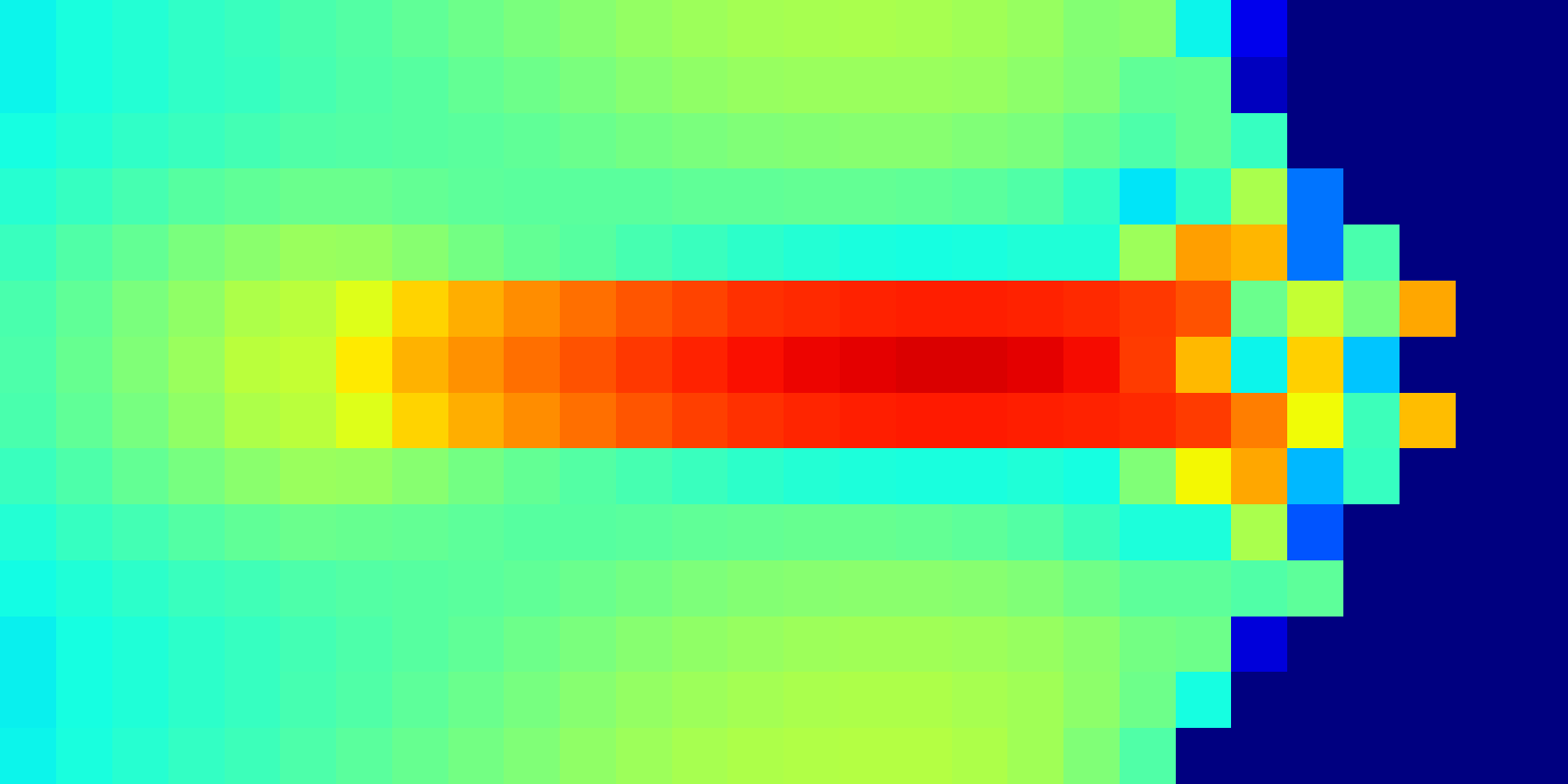} \\
\includegraphics[width=2.5in,keepaspectratio=true]{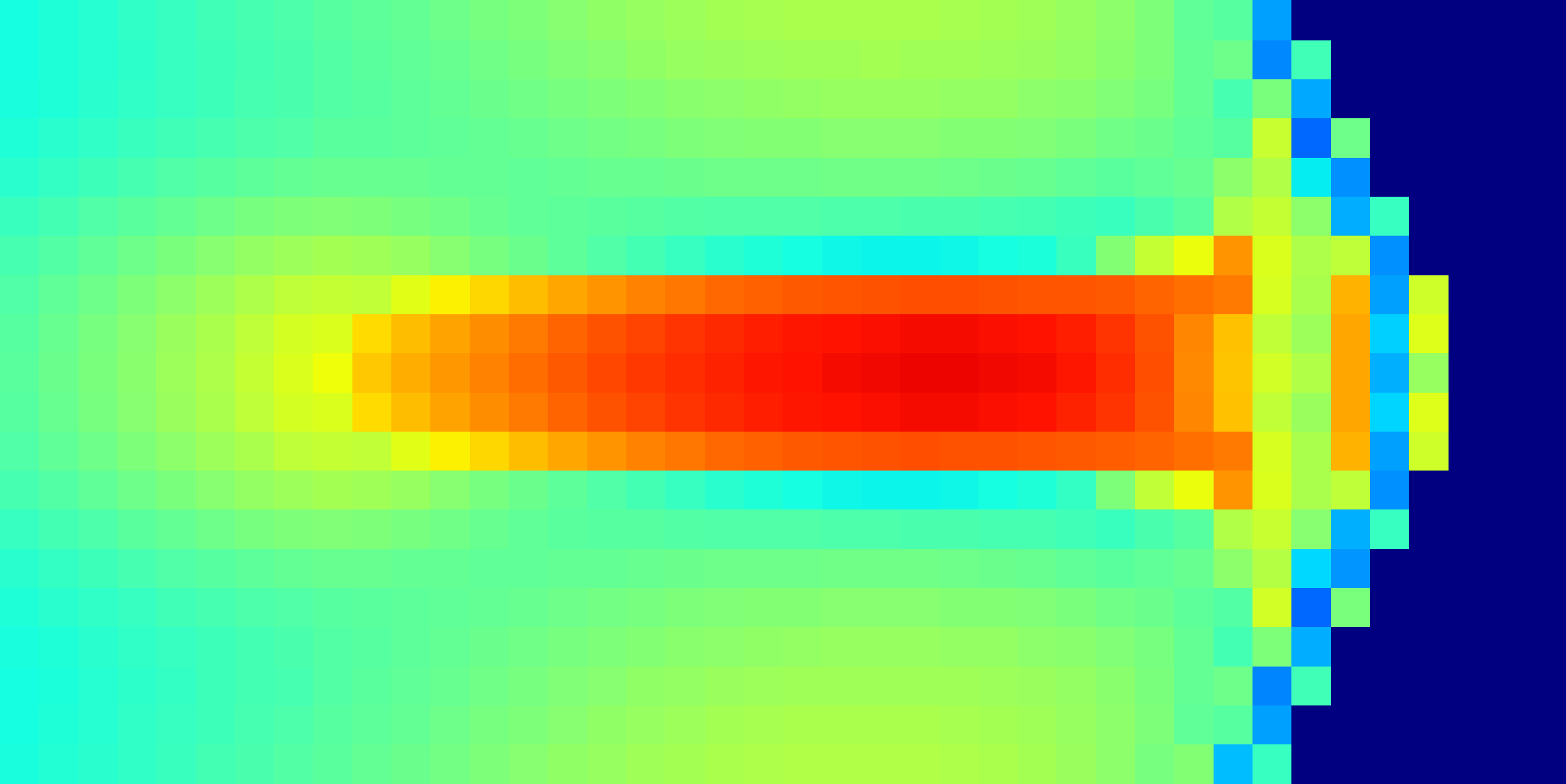} \\
\includegraphics[width=2.5in,keepaspectratio=true]{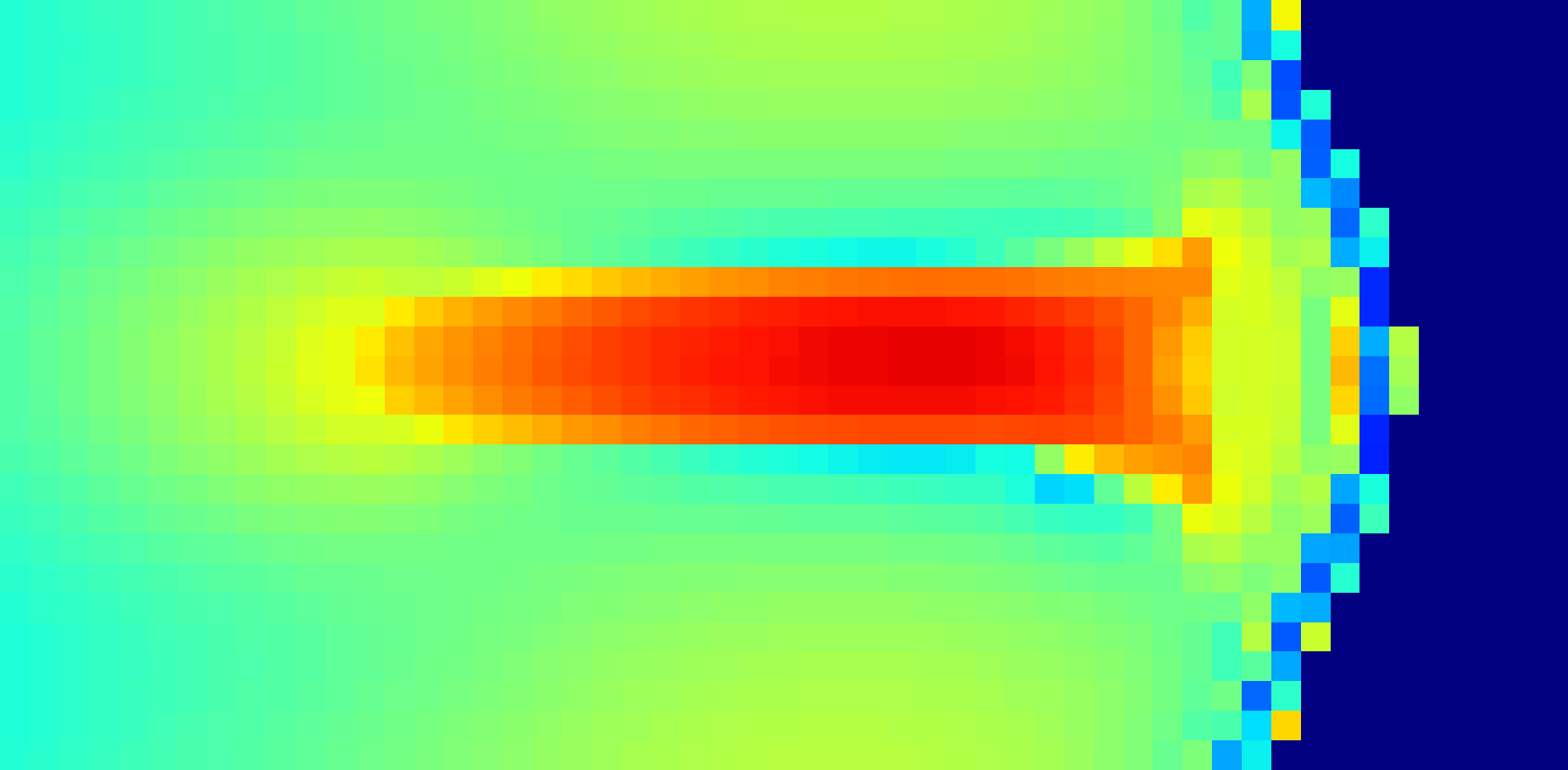} \\
\includegraphics[width=2.5in,keepaspectratio=true]{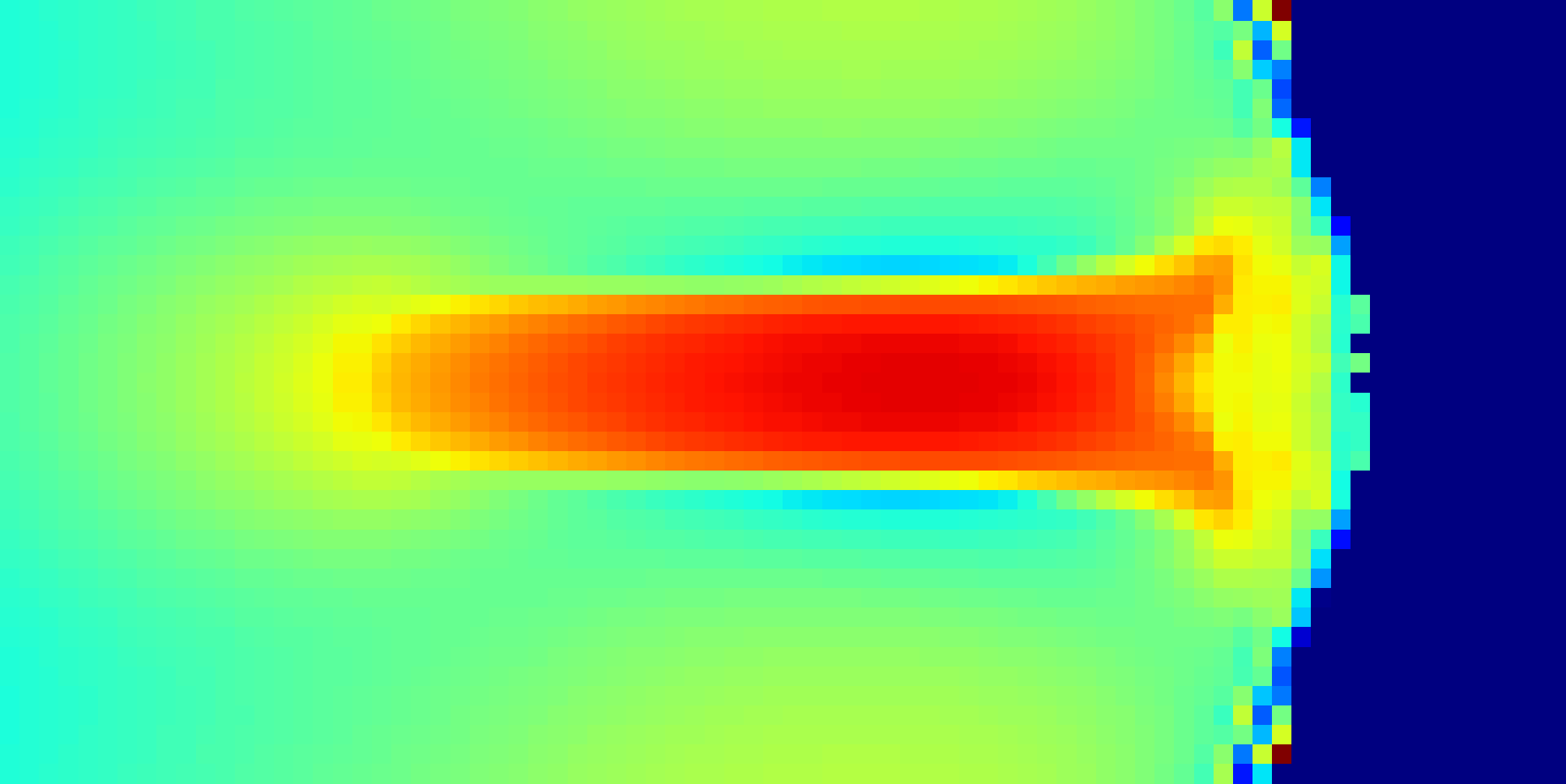}
\end{tabular}
}
\,
\includegraphics[height=3.5in,keepaspectratio=true]{colorbar_5km_csurf}
}
\caption{Detail of modeled horizontal surface ice speed (m/a) for experiment P1, in the same color scale used for Figure \ref{fig:PallSpeed}, at the end of 5 ka runs.  Horizontal resolutions (top to bottom): 15, 10, 7.5, and 5 km.}
\label{fig:P1speeddetail}
\end{figure}

\begin{figure}
\includegraphics[height=2.0in,keepaspectratio=true]{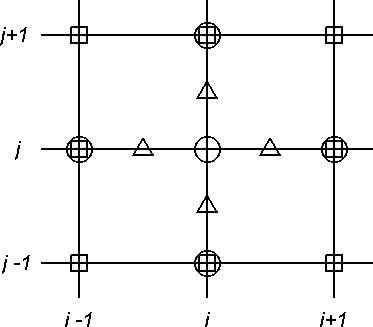}
\caption{A stencil for finite difference scheme \eqref{fdssaeqn1}, for approximating the first of Equations \eqref{SSA}.  Triangles show staggered grid points where $\bar \nu H$ is evaluated.  Circles and squares show where $v_1$ and $v_2$ are approximated, respectively.}
\label{fig:ssastencil}
\end{figure}

\begin{figure}[ht]
\includegraphics[width=3.0in,keepaspectratio=true]{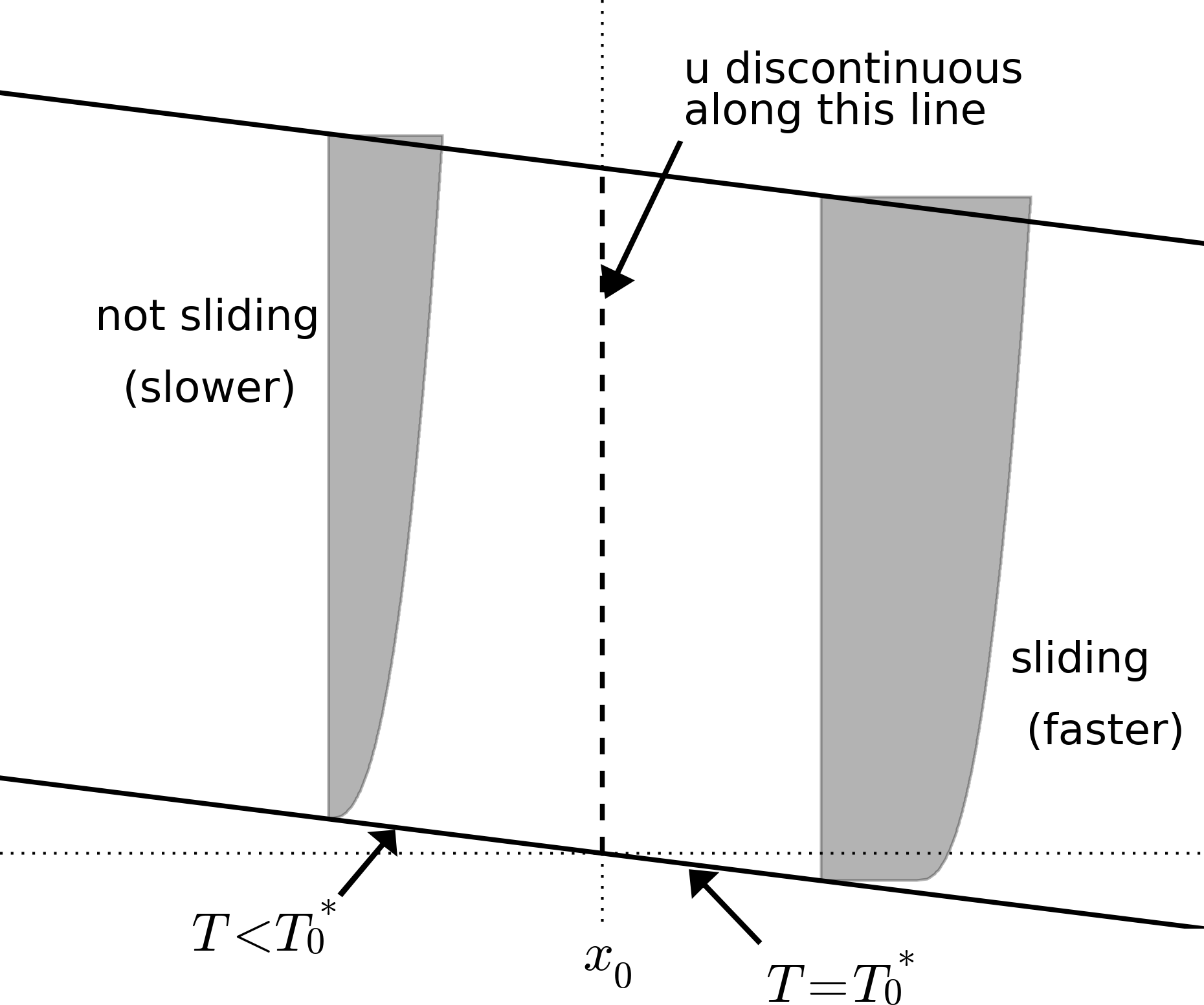}
\caption{If the sliding law ``turns on'' at a point where the pressure-melting temperature $T=T_0^*$ then there is a discontinuity in the sliding velocity everywhere in the ice column above that point.  The grey band suggests horizontal velocity $u$ as a function of elevation.}
\label{fig:slabjump}
\end{figure}

\end{document}